\documentclass[preprint,12pt,nonatbib]{elsarticle}

\usepackage{amssymb}
\usepackage{amsmath}
\usepackage[utf8]{inputenc}
\usepackage{tabularx}
\usepackage{lipsum}
\usepackage{array}
\usepackage{multirow}
\usepackage{booktabs} 
\usepackage{siunitx}  
\usepackage{comment}
\usepackage{placeins}
\usepackage{arydshln} 
\usepackage{hyperref}
\usepackage[backend=biber, sorting=none, url=false, doi=true]{biblatex}
\newcolumntype{Y}{>{\centering\arraybackslash}X}

\journal{Sciences Advances}

\begin{document}
    
\begin{frontmatter}



\title{Real-time vacancy concentration evolution revealed via heavy ion irradiation experiments} 


\affiliation[label1]{organization={MIT Plasma Science and Fusion Center (PSFC)},
             city={Cambridge},
             postcode={02139},
             state={MA},
             country={USA}}
\affiliation[label2]{organization={MIT Department of Materials Science \& Engineering},
             city={Cambridge},
             postcode={02139},
             state={MA},
             country={USA}}
\affiliation[label3]{organization={MIT Department of Nuclear Science \& Engineering},
             city={Cambridge},
             postcode={02139},
             state={MA},
             country={USA}}
\author[label1,label2]{Elena Botica-Artalejo}
\author[label1]{Gregory Wallace}
\author[label1,label3]{Michael P. Short}


\begin{abstract}
We show that \emph{in situ} ion irradiation transient grating spectroscopy ($I^3TGS$) can be used to monitor the real-time evolution of vacancy concentration generated by self-ion radiation damage in Cu-based alloys. Surface acoustic wave (SAW) frequencies are shown, using a combination of theory and experiment, to reveal vacancy concentrations and their kinetics in real-time. These results are shown to agree with corresponding atomistic kinetic Monte Carlo simulations at similar temperatures and dose rates. These results suggest the utility of TGS as a non-contact, non-destructive tool for real time defect monitoring.
\end{abstract}

\begin{graphicalabstract}
\includegraphics[width=1\textwidth]{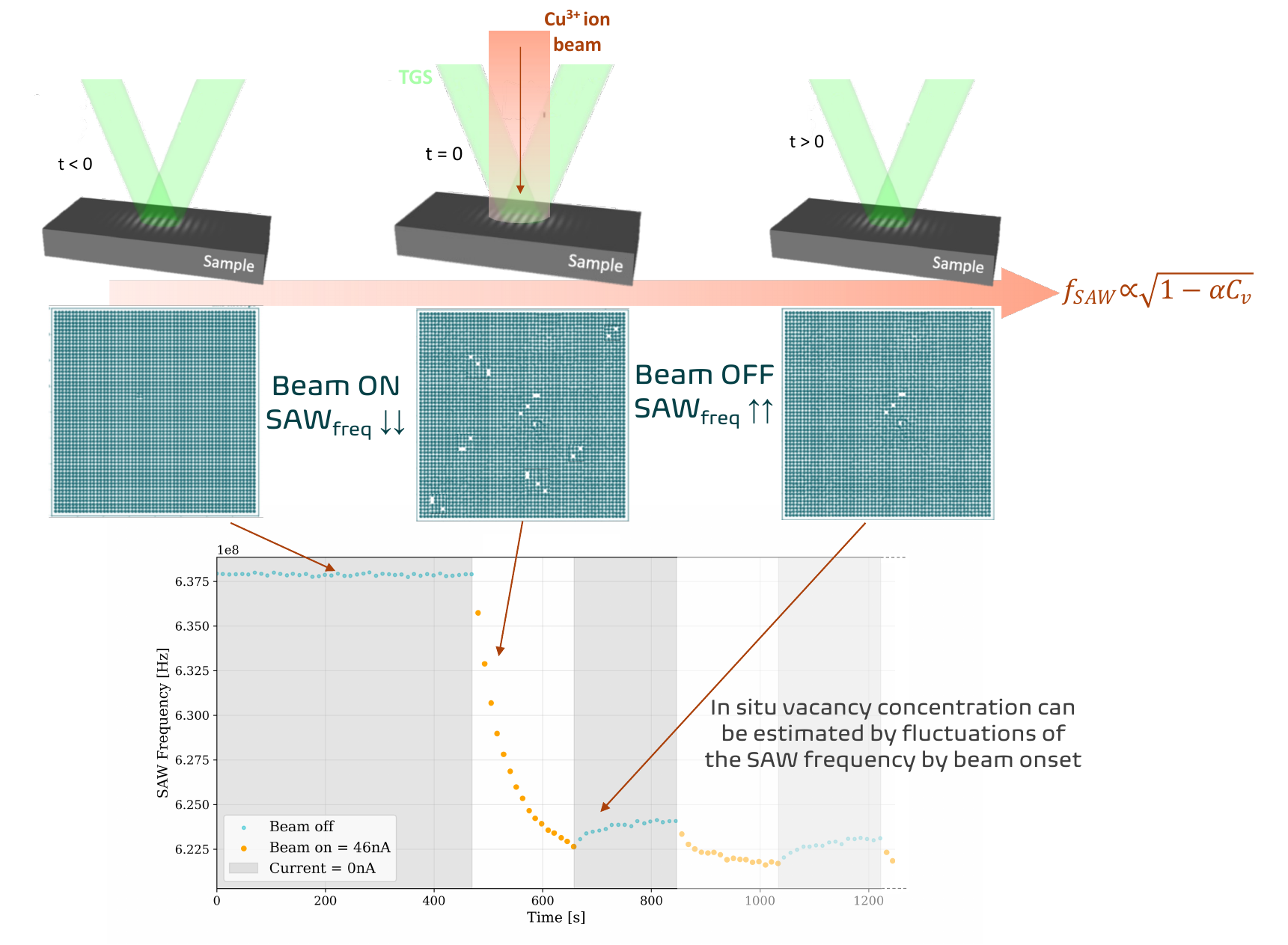}
\end{graphicalabstract}

\begin{highlights}
\item TGS surface acoustic waves directly monitor vacancy concentrations during irradiation
\item TGS-measured vacancy concentrations are shown to be independent of temperature
\item Kinetics of vacancy evolution during ion beam blanking can now be directly measured
\item \textit{In situ} TGS-measured vacancy concentrations agree with kinetic Monte Carlo and theory
\end{highlights}


\end{frontmatter}



\section{Introduction}

Fusion reactors pose a unique set of challenges for functional materials. All materials facing the plasma are subjected to severe neutron, ion, and heat fluxes, and even those components not directly exposed to the plasma will experience significant energetic neutron fluxes \cite{zinkle_designing_2014}. This neutron irradiation causes atoms to displace from their stable positions, generating microstructural damage, leading to property degradation through embrittlement, hardening, and void swelling, among others \cite{nordlund_primary_2018, zinkle_materials_2013}. A radiation damage resistant material is necessary to assure that material properties remain sufficiently stable over the lifetime of the fusion reactor. For example, tungsten, a candidate material for plasma facing components (PFCs), still presents noticeable microstructural degradation when is exposed to ion and helium implantation under operational conditions \cite{das_recent_2019}. Likewise, high-conductivity copper alloys, such as CuCrZr or GRCop-84, are planned to be used in high heat flux components, like monoblock pipelines or radio frequency (RF) antennas \cite{eldrup_influence_1998, wallace_multiphysics_2021}. GRCop-84, a precipitation-hardened Cu–8Cr–4Nb (at. \%) alloy with exceptionally high strength and thermal conductivity, has been proposed for fusion applications such as RF antennas \cite{seltzman_nuclear_2020}, but a more complete radiation damage resistance assessment is needed. In all cases, understanding radiation damage resistance requires tracing how the fundamental defects created by cascades evolve and affect material property deterioration. Only then, one can quantitatively judge which material is most resistant to the deleterious effects of radiation damage, by linking reduced defect concentrations and their dynamics to better-retained material properties.

When assessing radiation damage resistance in a material, multiple parameters may be considered. Each radiation damage cascade produces Frenkel pairs, vacancies and interstitials, some of which will remain in the lattice. These point defects are the root cause of later damage evolution. Vacancies and interstitials may migrate, recombine, or cluster into loops and voids, and their behavior dictates macroscopic changes. In fact, the basic reactions that control the steady-state defect concentration are precisely the recombination of these point defects and their interactions with interfaces or other microstructural features, colloquially known as damage sinks \cite{was_fundamentals_2017}. Usually, it is not possible to observe these point defects until they form clusters, at least in specimens large enough to represent bulk behavior (i.e. not ultra-thin transmission electron microscopy (TEM) foils). As a result, TEM can in some cases miss clusters that are smaller than the resolution limit \cite{hirst_revealing_2022}, or they may never evolve in a TEM foil due to its very small size and proximity of all defects to free surfaces. Moreover, most of these microstructural characterizations to assess damage are performed \emph{ex situ}, neglecting the dynamics of point defects and the real-time evolution of the microstructure. Additionally, the majority of them are destructive techniques that require thinning or sectioning of the material, altering the very microstructural features they aim to measure. Ultimately this provides only a single snapshot of the microstructure, preventing repeated measurements on the same sample. Some advanced facilities integrate \textit{in situ} TEM with ion beam irradiation, enabling direct observation of the formation and evolution of irradiation-induced defect structures, as well as insight into their kinetic behavior \cite{lu_enhancing_2016, kiritani_microstructure_1994, kaoumi_thermal_2008, taylor_situ_2017}. However, the construction and maintenance of such specialized facilities require substantial financial resources that are beyond the reach of most institutions, and access to beam time is typically granted only through competitive proposal-based allocations, which must be approved prior to use. Additionally, because some defects remain “hidden” to these techniques, experimental assessment of point defects is still a challenge, and non-conventional techniques that rely on other material properties that are affected by radiation damage need to be considered \cite{connick_measuring_2025}.

Although these defects are difficult to quantify experimentally, in principle computational models can provide a more complete picture of defect production and evolution \cite{gutierrez-camacho_towards_2025} in materials.  However, multi-scale challenges are present due to limitations of each computational method \cite{nordlund_primary_2018}. Density functional theory (DFT), is used mostly to obtain the interatomic potentials that will be used later in molecular dynamics (MD), or obtaining specific properties that can be used to calculate point defect evolution, like diffusivity or defect migration energetics \cite{dudarev_density_2013}. Classical MD with empirical potentials can handle keV-scale cascades and modest cell sizes \cite{jin_thermodynamic_2018,deluigi_simulations_2021}, but becomes remarkably expensive for high doses or large volumes, although the agreement with some irradiation quantities could be good if an appropriate interatomic potential is used. On the contrary, fast binary collision approximation (BCA) methods can simulate higher-energy events efficiently, yet they often underestimate the number of displaced atoms and fail to capture clustering correctly \cite{ortiz_combined_2018}. When BCA is combined with other techniques, much better predictions can be achieved, as is the case for BCA+MD \cite{gutierrez-camacho_towards_2025}. Combining different methods improves the reliability of the calculation, through the ability to mimic material behaviors at different time-scales, at the price of a higher computational cost. Even though computational methods can give insight into specific values, that could be experimentally unquantifiable, and help to estimate the evolution of the damage in the microstructure upon irradiation, the uncertainty about how reliable is that number due to the constraints imposed in the simulations is still a risk. Then, finding an experimental route to validate computational models and simulations for radiation damage is still needed.

Assessing radiation damage resistance is commonly carried out through void quantification and characterization using TEM imaging \cite{chen_-situ_2023, fan_diffusion-mediated_2021}, or computationally by evaluating the size and number of defects created after a given number of cascades \cite{jin_thermodynamic_2018}. Each approach carries inherent drawbacks, as discussed above, highlighting the need for a faster yet reliable assessment method. In this work, we explore transient grating spectroscopy (TGS) as a effective \emph{in situ} non-destructive tool to predict radiation damage resistance among materials manufactured under the same conditions.

In summary, accurately predicting radiation damage resistance boils down to understanding point-defect behavior, as point defect concentrations are the driving force terms in the differential equations of defect kinetics. In this work, \emph{in situ} irradiation coupled with TGS is used to assess the damage level in GRCop-84 and similar compositions at the same time that they are being irradiated, by means of the surface acoustic wave (SAW) frequency which is related with the material's elastic properties. First, we prove that the fluctuations seen on TGS data during ion beam blanking on and off are uniquely caused by radiation damage, successfully eliminating the possibilities of a trivial temperature effect. Then, the relation of the SAW frequency with the vacancy concentration in the material is demonstrated through a combination of theoretical prediction and experimental validation, to evaluate the increase in the vacancy concentration with increasing irradiation dose. Here we quantify radiation dose in displacements per atom (DPA) using the Norgett-Torrens-Robinson (NRT-DPA) method \cite{norgett_proposed_1975}. Finally, the ratio of TEM-measured void densities in two different samples is compared by the ratio of expected cumulative vacancy concentration, obtaining a strong agreement between both ratios, This suggests that TGS is a promising technique to estimate \emph{in situ} radiation damage resistance in materials via its most fundamental and atomic scale property - point defect concentrations as the root of all damage, as all other defects (dislocation loops, clusters, voids, precipitates) evolve directly from them.

\section{Methods}
\label{sec2}

\subsection{Material preparation}
\label{subsec1}
CuCrNb and CuCrTa combinatorial thick films were created using physical vapor deposition (PVD) in a Kurt J. Lesker PVD 75 PRO Line using magnetron sputtering. The three different elements that compose the alloy were simultaneously deposited, using specific deposition parameters for each material (see Table \ref{tab:deposition_parameters}), to obtain Cu-rich samples. Spatial variations laid down by each sputtering target creating a sample with compositional gradients in two dimensions across the substrate. Targets were placed at equal (or almost equal) angular spacings from each other to ensure gradient homogeneity, so just the gun power is the contributing factor for material distribution along the wafer substrate. Additionally, the same inclination with respect to the substrate was kept for all sputtering guns. Different batches were done in order to explore different compositional spaces by target power variations. The targets used from the Kurt J. Lesker Company include Cu (99.99\%), Cr (99.95\%), Ta (99.95\%) and Nb (99.95\%). Impurity certificates can be found in our data repository for this manuscript \cite{botica-artalejo_repository_2025}. The material used as a substrate was 1-inch p-type (B-doped) silicon wafer polished on one side to semi-Prime quality, with a R$_a$ < 10\AA. To increase adhesion to the substrate and decrease interfacial stresses due to lattice mismatch, a layer of Ti (99.995\%) was deposited before deposition of the thick film. Samples with a film thickness between $4 - 5\mu m$ were obtained, specifically to grow films far beyond the epitaxial thin film limit where substrate strain dominates microstructure. Therefore, films were grown thick enough such that more bulk-like microstructures were obtained. The thickness varied across the sample, with thinner regions on the outer diameter of the PVD substrate and thicker in the center. The variation of composition inside a wafer changes from 1-2 at.\% from side to side, depending on the element considered.

\begin{table}[h]
\centering
\small
\begin{tabularx}{\textwidth}{>{\raggedright\arraybackslash}X c c c c}
\toprule
\textbf{Parameter} & \textbf{Batch 1} & \textbf{Batch 2} & \textbf{Batch 3} & \textbf{Batch 4} \\ 
\midrule
Vacuum level prior to deposition (Torr) & \(8.7 \times 10^{-6}\) & \(8.1 \times 10^{-6}\) & \(7.4 \times 10^{-6}\) & \(7.4 \times 10^{-6}\)\\ 
Chamber pressure at deposition (Torr)   & \(5.9 \times 10^{-4}\) & \(4.4 \times 10^{-4}\) & \(6.2 \times 10^{-4}\) & \(6.2 \times 10^{-4}\) \\ 
Sputtering time (min)                   & 150 & 140 & 130 & 150\\ 
Gas flow (sccm)                         & 12  & 33  & 20  & 20\\ 
Gas species                             & \multicolumn{4}{c}{Argon (99.999\%)} \\ 
Temperature (\(^\circ\)C)               & \multicolumn{4}{c}{20--25} \\ 
Cu target power \& size (Watts/inch/type)    & 150/1/DC & 160/1/DC & 160/1/DC & 150/2/DC \\ 
Cr target power \& size (Watts/inch/type)    & 45/2/DC  & 40/2/RF  & 40/2/RF  & 80/2/RF  \\ 
Nb target power \& size (Watts/inch/type)    & 35/2/DC  & 30/2/DC  & 35/2/DC  & -  \\ 
Ta target power \& size (Watts/inch/type)    & -  & -  & -  & 30/2/DC  \\
Ti target power \& size (Watts/inch/type)    & \multicolumn{4}{c}{60/2/DC} \\ 
Ti sputtering time (min)                & \multicolumn{4}{c}{60} \\ 
\bottomrule
\end{tabularx}
\caption{Deposition parameters for CuCrNb and CuCrTa thick films.}
\label{tab:deposition_parameters}
\end{table}

\subsection{Microscopy characterization}
To obtain the elemental composition on each of the wafers, a Zeiss Gemini 450 scanning electron microscope (SEM) with an energy-dispersive X-ray spectroscopy (EDS) detector was used with an energy of 17 kV and a probe current of 1 nA, to excite the characteristic X-ray lines while keeping a reasonable interaction volume. The working distance was set to 8.5 mm as it optimizes X-ray count rates in this instrument. In order to assess the relative amount of each element in the sample and neglect contamination, carbon and oxygen energy peaks have been deconvoluted in the EDS analysis. Compositions and the type of test performed on each sample are shown in Table \ref{tab:chem_comp}. 

\begin{table}[h]
    \centering
    \small
    \setlength{\tabcolsep}{3pt}
    \begin{tabular}{@{}l c c c c c >{\centering\arraybackslash}m{2.6cm}@{}}
    \toprule
    \textbf{SampleID} & \textbf{Batch} & \textbf{Cu (at.\%)} & \textbf{Cr (at.\%)} & \textbf{Nb (at.\%)} & \textbf{Ta (at.\%)} & {\textbf{Experiment type}} \\
    \midrule
    031\_0101 & 1 & 68.4 & 25.3 & 6.4 & -- & \multirow{4}{*}{\parbox{2.6cm}{\centering Beam on/off Heavy ions vs. Protons}} \\
    031\_0301 & 1 & 78.7 & 11.8 & 9.5 & -- & \\
    031\_0305 & 1 & 90.1 & 6.1 & 3.9 & -- & \\
    031\_0404 & 1 & 87.3 & 7.3 & 5.4 & -- & \\
    \midrule
    042\_10 & 2 & 66.9 & 24.5 & 8.5 & -- & \parbox{2.6cm}{\centering Pulsed beam + DPA log. scale} \\ 
    \midrule
    045\_04 & 3 & 93.2 & 4.5 & 2.2 & -- & \multirow{3}{*}{\parbox{2.6cm}{\centering Beam on/off in DPA log. scale}} \\ 
    044\_15 & 4 & 86.9 & 9.2 & -- & 3.8 & \\
    044\_18 & 4 & 70.4 & 24.3 & -- & 5.3 & \\ 
    \bottomrule
    \end{tabular}
    \label{tab:chem_comp}
    \caption{Chemical composition and corresponding batch for CuCrNb and CuCrTa thick films.}
\end{table}

To perform microstructural characterization of the samples after irradiation, transmission electron microscopy (TEM) was done with a JEOL F200. This microscope was used at 200kV, in conventional TEM mode using the double-tilt holder and a objective aperture of $30\mu m$, which helped to decrease contrast to better visualize voids. To measure void size, Fiji (ImageJ 1.54p) software was used. Voids were counted and measured manually using the ROI Manager tool, as the diverse contrast in the images didn't allowed for automatization of the process. Then, the total imaged area was calculated to obtain the void areal density. To prepare the TEM lamellae, focused-ion beam (FIB) removal of specimens was performed using the FEI Helios 660. Samples were milled down to $40-50\mu m$ in thickness by periodically decreasing the ion current from 30kV to 2kV. Finally, ion mill cleaning was performed before TEM inspection to remove any Ga damage and oxides from the surface, as copper can easily oxidize. This was done using a Fischione 1051 TEM mill.

\subsection{\textit{In situ} ion irradiation transient grating spectroscopy ($I^3TGS$)}
\begin{figure}
    \centering
    \includegraphics[width=0.9\textwidth]{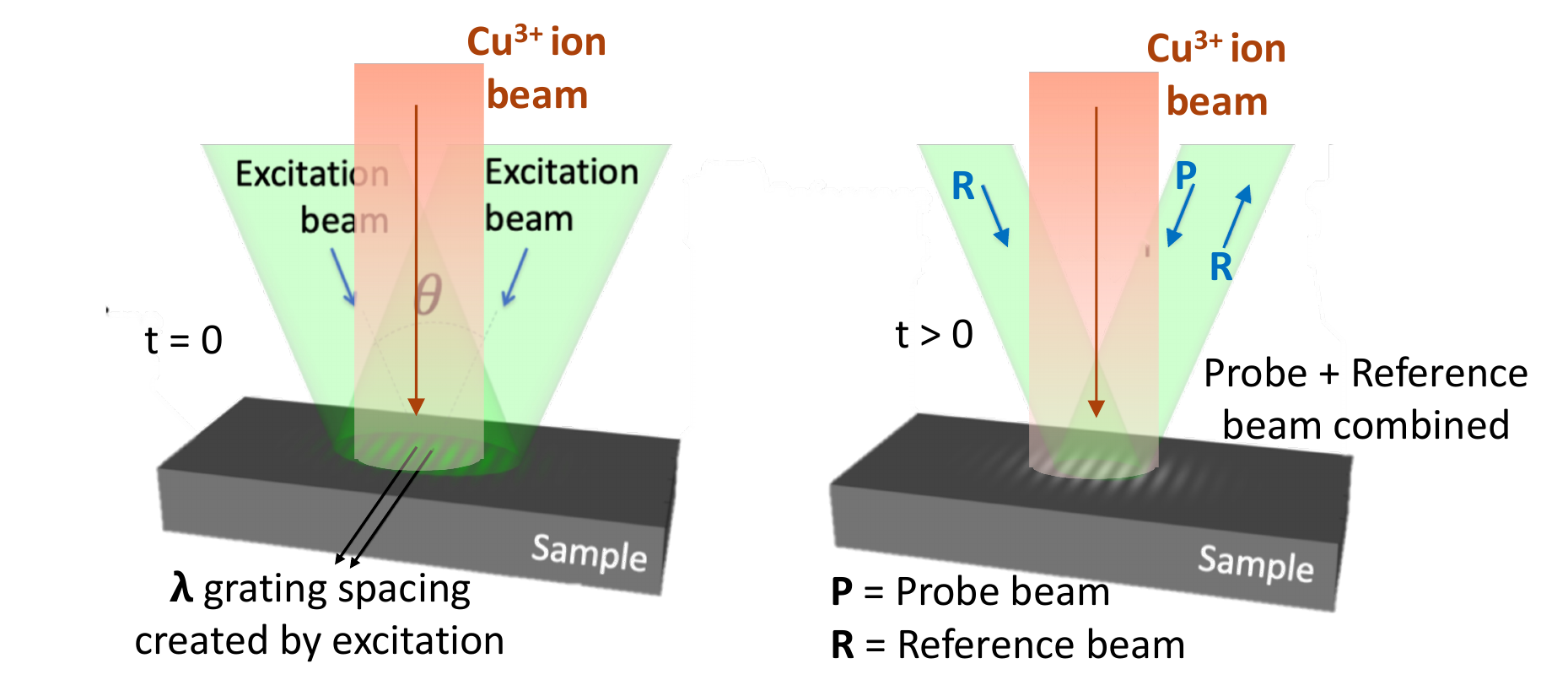}
    \caption{Scheme of TGS coupled with \emph{in situ} irradiation. Adapted from \cite{hofmann_transient_2019}.}
    \label{fig:TGS_phy}
\end{figure}

Transient grating spectroscopy (TGS) is a non-contact, non-destructive laser technique that facilitates the quantitative evaluation of a material’s elastic properties and thermal diffusivity at controlled depths \cite{hofmann_transient_2019}. This technique can provide localized measurements of the thermo-elastic response of the material, with a spatial resolution of the focal width of the probe laser employed. A conceptual representation of this technique coupled with ion beam irradiation can be seen on Figure \ref{fig:TGS_phy}.
A first beam, consisting of a $1kHz$, $532nm$, Q-switched laser with a pulse length around $500ps$, will excite the material surface creating a grating pattern with an on-sample power of $7.36mW$. This is called the "pump beam", as it provides the excitation energy to launch surfae acoustic waves (SAWs) for measurement. The grating spacing will determine the the maximum depth from which thermo-elastic properties can be extracted. Then, a continuous wave probe $577nm$ single longitudinal mode laser will monitor this surface excitation via the first-order diffraction of the quasi-continuous wave with an intensity around $127kW m^{-2}$ and a power on sample of $4.6mW$. This is called the "probe beam", as its time-dependent diffraction contrast constitutes the TGS signal actually measured. A detailed description of all TGS parameters can be found in Table \ref{tab:tgs_param}. The probe beam directly measures the SAW frequency revealing elastic properties as well as thermal diffusivity by the decay of the signal, $I_{tot}$, which has the shape of:
\begin{equation}
I_{tot} (t)= A \Big[ 
\mathrm{erfc} \Big( q_0 \sqrt{\alpha_x t} + \frac{\beta}{\sqrt{t}} e^{-q_0^2 \alpha_x t} \Big)
\Big]
+ B \Big[ \sin(2 \pi f_{SAW} t + \Theta) e^{-t/\tau} \Big] + C
\end{equation}
where $q_0 = 2\pi\Lambda$ is the grating wavenumber resulting from a TGS grating spacing of $\Lambda [m]$ (the focusing length configuration on the TGS device used here is 1:1 through the phase mask, then $\Lambda$ is employed indistinguishable from $\lambda$, the wavelength); $\alpha_x [m^2s^{-1}]$ is the surface-plane thermal diffusivity; $\beta [s^{1/2}]$ describes the ratio of contributions from displacement to reflectivity; $f_{SAW}[Hz]$ is the surface acoustic wave (SAW) frequency, $\Theta$ the phase, and $\tau [s]$ decay time of the surface acoustic response \cite{dennett_thermal_2018}. Finally \(A\), \(B\), and \(C\) are constants to fit using the experimental data to a  non-linear model \cite{wylie_accelerating_2025}. 

The bridge that will relate this signal with the elastic properties of the material will be the Rayleigh wave equation \cite{royer_rayleigh_1984}: 
\begin{equation}
c_{r} = f_{SAW} \lambda = \sqrt{\frac{E}{\rho}} \, 
\end{equation}
where $\lambda [m]$ is the wavelength of the spatially periodic intensity pattern generated on the material surface, $E [Pa]$ the Young's modulus and $\rho [kg/m^3]$ is the material density.
For the case of an elastically isotropic material or a polycrystalline material with small enough grains that it presents as elastically isotropic to the TGS beam spots, this approximation will give quantitative results. Then, the SAW frequency can be rewritten in terms of the elastic modulus of the material as follows:

\begin{equation}
f_{SAW} = \frac{1}{\lambda} \sqrt{\frac{E}{\rho}}
\label{eq:SAW_to_E}
\end{equation}

However, for elastically anisotropic materials, which is the case of most polycrystalline materials due to the spot-size of this technique in comparison to the average grain-size, the approximation is not direct, so a better approach is using relative values rather than absolute ones. 

\begin{figure}
    \centering
    \includegraphics[width=0.85\textwidth]{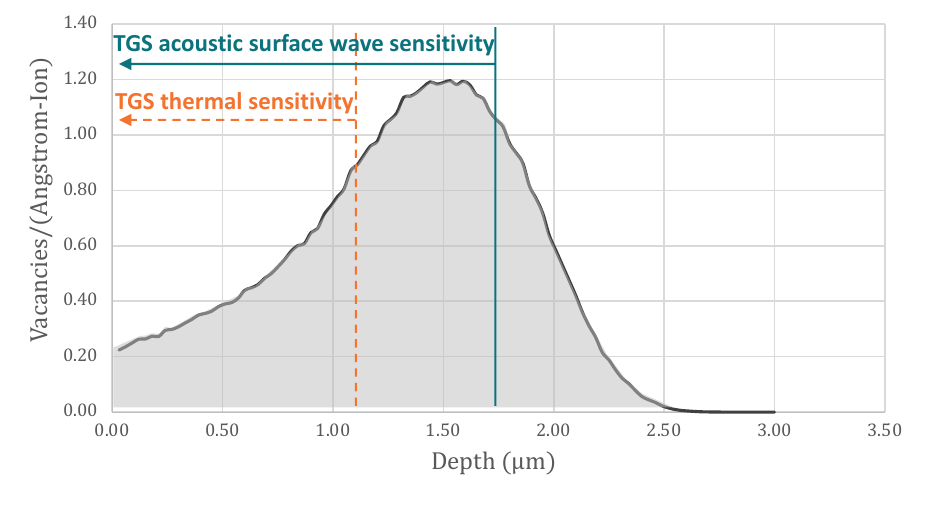}
    \caption{Irradiation depth profile as calculated by SRIM and nominal TGS e-folding depth sensitivity for thermal and acoustic wave properties.}
    \label{fig:TGS_ion_damage}
\end{figure}

To decrease the on-sample power, a neutral density filter of 0.2\% was applied to the pump laser, reducing the pump on sample power to $5.89 mW$. The setup used was with a di-homodyne phase collection geometry, as the one described in reference \cite{dennett_thermal_2018}. This type of setup uses two probe lasers that allow the signal to be more stable against small changes in the laser phase, after both being fixed to $\phi = \pm \tfrac{\pi}{2}$. The current setup employed in these experiments is shown in \cite{wylie_accelerating_2025}. One of the advantages of this technique is that the probe depth can be adjusted, just by changing the grating spacing in the system. This tunes the nominal sensitivity for thermal properties and for the SAW signal, which are $\lambda/\pi$ \cite{kading_transient_1995}, and $\lambda/2$ \cite{royer_rayleigh_1984} respectively, to the depth of the ion irradiation. A nominal value of $\Lambda =3.4 \mu m$ was chosen to match the range of the damaged region by the irradiating ions (see Figure \ref{fig:TGS_ion_damage}). Prior to TGS experiments performed each day, a calibration of the current optical setup was performed using a single crystal tungsten sample with a \{100\} surface orientation in order to determine the real grating spacing, as it could have slight variations ($< \pm 0.1 \mu m$) due to volumetric expansion caused by ambient conditions.

In order to increase the time resolution of the technique, measurements with different numbers of traces averaged per cycle were tested, from 10000 traces in 1 minute up to 1000 traces in 7 seconds. In the end we increased the time resolution by 6 without decreasing the quality of the data, by averaging 2000 signals every 10 seconds. To process the TGS raw data and obtain the values of the SAW frequency and thermal diffusivity, a postprocessing MATLAB code was employed \cite{short_lab_github_2025}. TGS-processed and raw data can be found in the open data repository for this article.

\begin{table}[h]
    \centering
    \begin{tabular}{c c}
    \toprule
        \textbf{Parameter} & \textbf{Value} \\
        \hline  
        Pump wavelength & 532 nm \\
        Pump energy per pulse & 10 $\mu$J \\
        Pump pulse duration & $<$500 ps \\
        Pump spot size (on-sample) & 150 $\mu$m \\
        Pump firing frequency & 1 kHz \\
        Pump on-sample power & 7.36 mW \\
        Probe wavelength & 671 nm \\
        Probe frequency & Continuous Wave \\
        Probe spot size & 90 $\mu$m \\
        Probe on-sample power & 4.6 mW \\
        Detector bandwidth & 50 kHz - 1 GHz (3 dB) \\
        Oscilloscope bandwidth & 5 GHz \\
        Averaged traces per second & 2000 \\
        Calibrated on-sample grating spacing & 3.4 $\mu$m \\
        Sample roughness& $R_a = 10$ \AA \\
        \hline
    \end{tabular}
    \caption{Detailed summary of TGS system parameters. Spot sizes reported as $2\sigma$ Gaussian widths.}
    \label{tab:tgs_param}
\end{table}

TGS was coupled to one of the chamber viewports at the end of an ion accelerator beamline, allowing one to monitor the evolution of the thermo-elastic properties during irradiation. Because the TGS needs to be completely perpendicular to the sample in order to satisfy the conditions for the homoodyne signal to form, the sample was rotated 45$^\circ$ with respect the incident ion beam, so ions were implanted at that angle.

Samples were self-ion irradiated at the Cambridge Laboratory for Accelerator Surface Science (CLASS) at the MIT Plasma Science and Fusion Center, a tandem ion accelerator of 1.7 MV of terminal potential. During the experiments Cu$^{3+}$ ions at 6.5-7 MeV and H$^+$ ions at 1.624 MeV were used. All irradiations were conducted at room temperature, without using any external heating source. Depending on the nature of the experiment different particle flux rates varied from $9.4\times 10^{15}$ - $7.0\times 10^{16} ions/m^2\cdot s$. The beam current was periodically monitored (at least every 15 min) throughout the irradiation using a Faraday cup. To focus the beam, an aperture of 2 mm diameter was employed, however, due to the 45$^\circ$ angle respect to the normal of the sample, the beam area was 4.44 $mm^2$ resulting in an ellipsoidal spot. Prior to the experiment, the beam was aligned with the TGS laser spot by performing a calibration using a thin ceramic layer coated with silver-activated zinc sulfide covering half of the sample, which allowed visualization of the beam position via ion beam luminescence. The irradiation penetration depth was estimated using the Stopping Range of Ions in Matter (SRIM) \cite{ziegler_srim_2010} software following the method described by Stoller et al. \cite{stoller_use_2013}. The default displacement energy was modified, using 30 eV for Cu, 40 eV for Cr, 60 eV for Nb and 90 eV for Ta \cite{astm_astm_2023}. On Figure \ref{fig:TGS_ion_damage}, a damage profile for GRCop-84 under the irradiation conditions represented in Figure \ref{fig:TGS_beam_comparison}.a) can be seen plotted on top of the contribution to the elastic and thermal properties from the TGS.

\subsection{Infrared camera measurements of sample temperature}
To monitor any potential ion beam heating, a Telops model FAST M3k infrared (IR) camera with a spectral range of 1.5$\mu m$ to 5.4$\mu m$ was used. The image was sent to the camera using a $45^\circ$ gold-coated mirror placed perpendicular to the material surface to avoid any angle-distortion of the image. The data was processed using a MATLAB script \cite{botica-artalejo_repository_2025}, obtaining the increase in temperature due to the ion beam. Prior to taking the measurements, a calibration needs to be done by heating the sample up to $100^\circ C$ in $5^\circ C$ increments.

\section{Results}
\subsection{\emph{In situ} TGS coupled with ion irradiation}
\begin{figure}
    \centering
    \includegraphics[width=\textwidth]{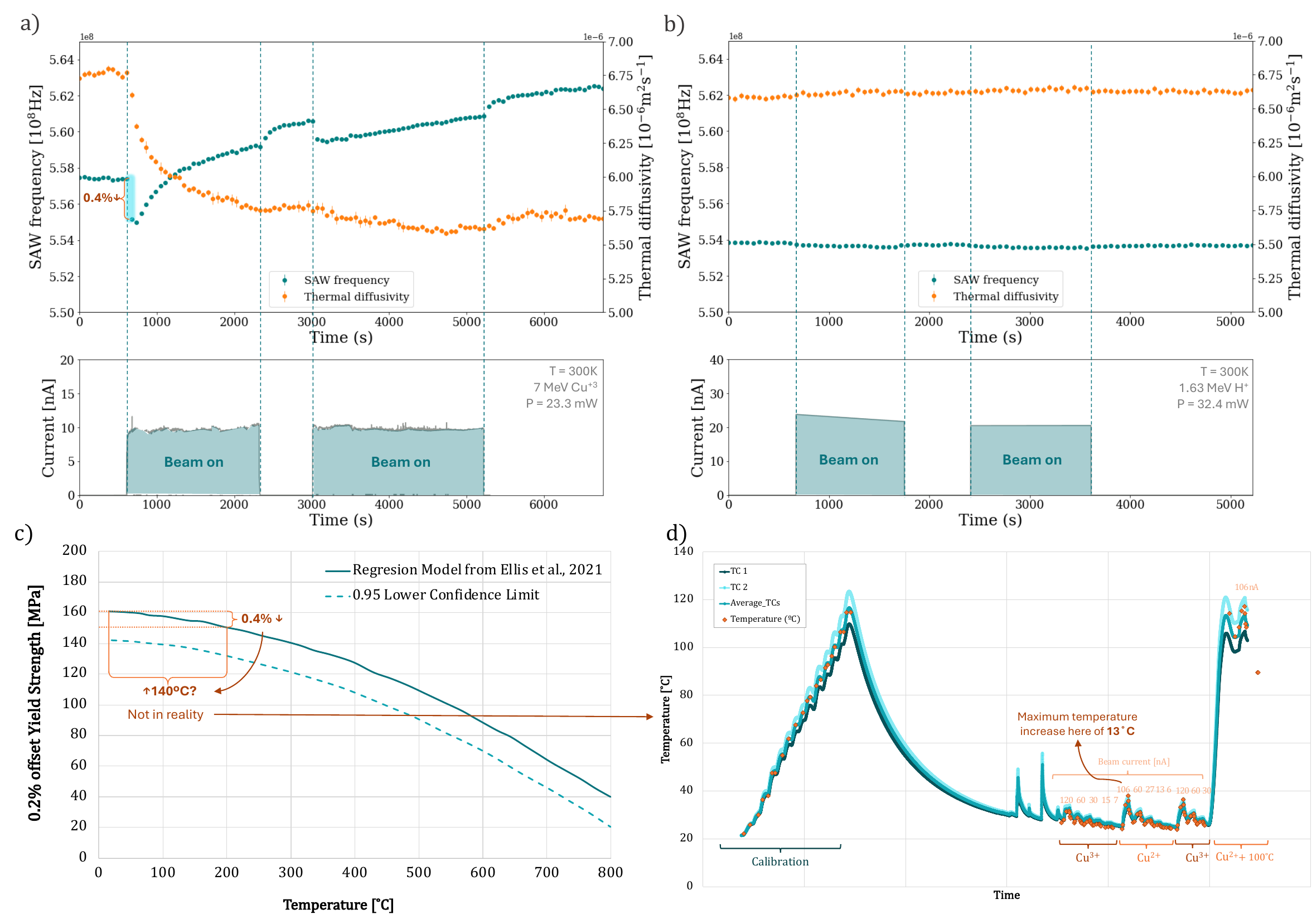}
    \caption{\emph{in situ} TGS results from 7 MeV Cu$^{3+}$ irradiation at 23.3 mW (a) and from 1.63 MeV proton irradiation (b). Most crucially, these two results show that two experiments with the same beam heating power in Watts produce changes in SAW frequency proportional to their ion stopping powers, and not to beam heating temperature. This isolates the effect of beam heating from vacancy concentration on the changes in SAW frequency and thermal diffusivity. c) 0.2\% offset Yield Strength model for GRCop-84 from \cite{ellis_aerospace_2001}, showing that a 0.4\% change in elastic properties would imply an increase in temperature of $140^\circ C$. d) Infrared camera calibration and temperature measurement for different isotopes of copper at 1.68MV terminal potential. This plot shows that the maximum increase in temperature for $Cu^{2+}$ is $13^\circ C$, and just around $1.5^\circ C$ for 15 nA of $Cu^{3+}$ (current conditions used in plots a) and b).}
    \label{fig:TGS_beam_comparison}
\end{figure}

\textbf{I}n situ \textbf{I}on \textbf{I}rradiation experiments coupled with \textbf{TGS} measurements (I$^3$TGS) were conducted on four individual samples from Batch 1 with different compositions, one of them being GRCop-84 \cite{ellis_grcop-84_2005} (the chemical composition of each sample can be found in Table \ref{tab:chem_comp}). During irradiation, the ion beam was deliberately interrupted twice, once after 18 minutes and once after 1 hour (corresponding to approximately half of the total irradiation duration). Figure \ref{fig:TGS_beam_comparison} shows the evolution of SAW frequency and thermal diffusivity with increasing dose under 7 MeV Cu$^{3+}$ (a)) and under 1.63 MeV H$^+$ (b)) irradiation for the sample with the GRCop-84 composition. The remaining compositions of Batch 1 exhibited the same SAW frequency behavior upon beam on/off cycling and different ions (figures available in the data repository from this paper). It is also important to note that the errors associated with the TGS measurements are quite small in all the samples, for the SAW frequency they are essentially imperceptible (less than 0.2 \%), and for the thermal diffusivity the average error is only about $\sim 0.4\%$. This low level of uncertainty is related to the small, controlled surface roughness ($R_a$) \cite{ruiz_laser-uitrasonic_2004} obtained by our chosen manufacturing method, magnetron-sputtering PVD, which produces compact films with uniform, smooth surfaces when a perfectly flat substrate is used. In our case a prime-quality Si wafer was employed as a substrate, with an $R_a\sim0.5nm$. 

Pulsed beam irradiations using $Cu^{3+}$ ions were repeated on samples from different PVD batches (2 and 3) to exclude any sputtering deposition effect associated to a specific batch. Beam interruptions were incorporated to verify that the response persists throughout irradiation and to track the SAW frequency and thermal diffusivity evolution with increasing DPA. 
As shown in Figure \ref{fig:TGS_beam_comparison}.a), each beam-on event causes a drop in SAW frequency, followed by recovery to a higher value when the beam is turned off.
These variations of the SAW frequency occur concurrently with the temperature variations imposed by the beam on the sample. The temperature increase induced by the ion beam should be analyzed separately, as most materials such as GRCop-84 become more compliant at higher temperatures. Therefore, this effect must be quantified and separated from any potential vacancy concentration effects explicitly.

\subsection{On-sample temperature by IR measurements}
To quantify the real sample temperature increase due to the beam, infrared camera measurements where performed for different sample materials, beam currents, Cu ions and accelerator terminal powers. Figure \ref{fig:TGS_beam_comparison}.d) shows one of those experiments, where the change in temperature for the condition of 106 nA of current, using Cu$^{2+}$ ions at 7 MeV is just $13^\circ C$, for the most extreme case investigated in this study. For the case equivalent to the conditions of Figure \ref{fig:TGS_beam_comparison}.a), b) the temperature increase is just $1.5^\circ C$, instead of the $140^\circ C$ expected from the evolution of elastic properties with temperature from \cite{ellis_aerospace_2001}. This sets an upper bound on any temperature effects causing SAW frequency to decrease or increase as the ion beam is turned on or off, respectively, allowing us to quantify the remaining changes as due to the presence of open volume defects. Then, because all open volume defects are comprised of the number of vacancies times the atomic volume to first order, we can extract the vacancy concentration as the sum of lone point defects and those bound in vacancy clusters.

\subsection{Periodic pulsed-beam experiments}
Following the reasoning from the model developed in section \ref{sec:mod_dev} to calculate the vacancy concentration from the beam on/off fluctuations in the SAW frequency, and considering the results from the atomistic kinetic Monte Carlo (akMC) simulations of E. Martinez et al. \cite{martinez_point_2020} of a very similar situation, a new experiment has been designed. In this experiment, performed at room temperature using $Cu^{3+}$ at 6.5 MeV and a dose rate of $1.7\times10^{-3}$ DPA/s, we pulse the ion beam on and off every three minutes with the purpose of observing both stable vacancy concentrations and their kinetics as a function of applied dose. As a result, the vacancy concentration in the material evolves according to the vacancies that are created and annihilated before the concentration of point defect reaches steady state. Then other point defect-related phenomena, like clustering or dislocation loop evolution, do not have enough time to develop to a significant degree. In addition, as mentioned before small vacancy clusters behave elastically like groupings of lone vacancies to first order, thus we should still semi-quantitatively capture both kinetics and total vacancy concentration evolution. Therefore, the changes that are observed in the SAW frequency can be attributed purely to fluctuations in the vacancy concentration induced by primary radiation damage, as they will be more dominant than interstitials (more detail in section \ref{sec:mod_dev_cas}). As a consequence, following the procedure described in section \ref{mod_dev_ev}, the vacancy concentration can be predicted in real time as radiation damage evolves. It is important to note that the dose rates used in these experiments are not directly comparable to those expected in future fusion reactors, where damage levels are projected to reach on the order of tens of DPA per year rather than several DPA per hour. Nevertheless, the higher dose rates employed during these experiments were intentionally selected to enable rapid and conservative assessment of the radiation damage resistance of the materials tested. Our aim was to demonstrate viability of the technique, not correspondence between ion irradiation and fusion reactors over four orders of magnitude in dose rate.

\begin{figure}[h]
    \centering
    \includegraphics[width=\textwidth]{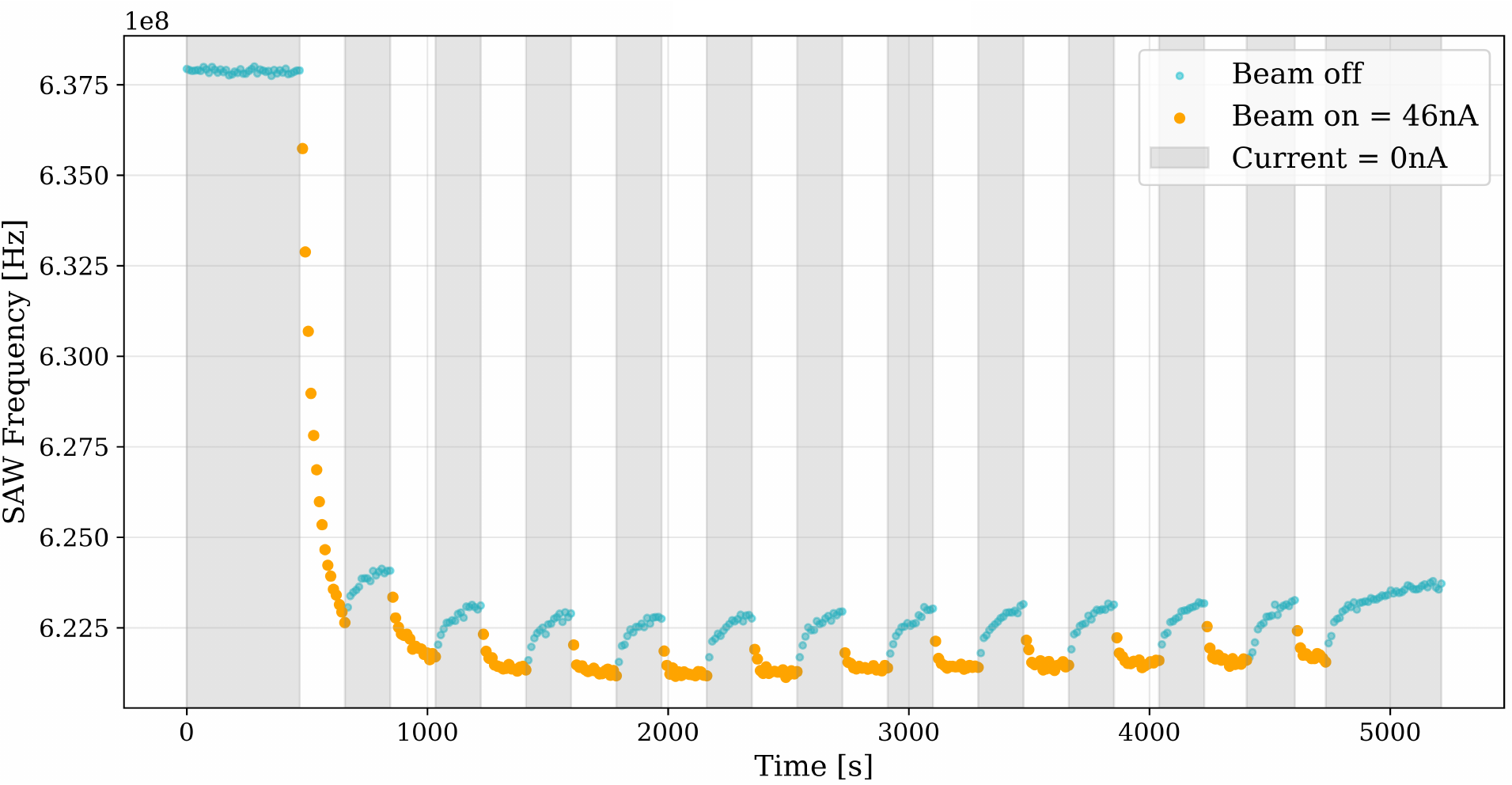}
    \caption{Periodic pulsed beam experiment, showing how SAW frequency evolves as the Cu-ion beam is cycled on and off. The white background and orange dots represent the periods in which the beam is on, and the grey background and blue dots represent the periods in which the beam is off.}
    \label{fig:Pulsed_beam}
\end{figure}

\begin{figure}[h]
    \centering
    \includegraphics[width=.7\textwidth]{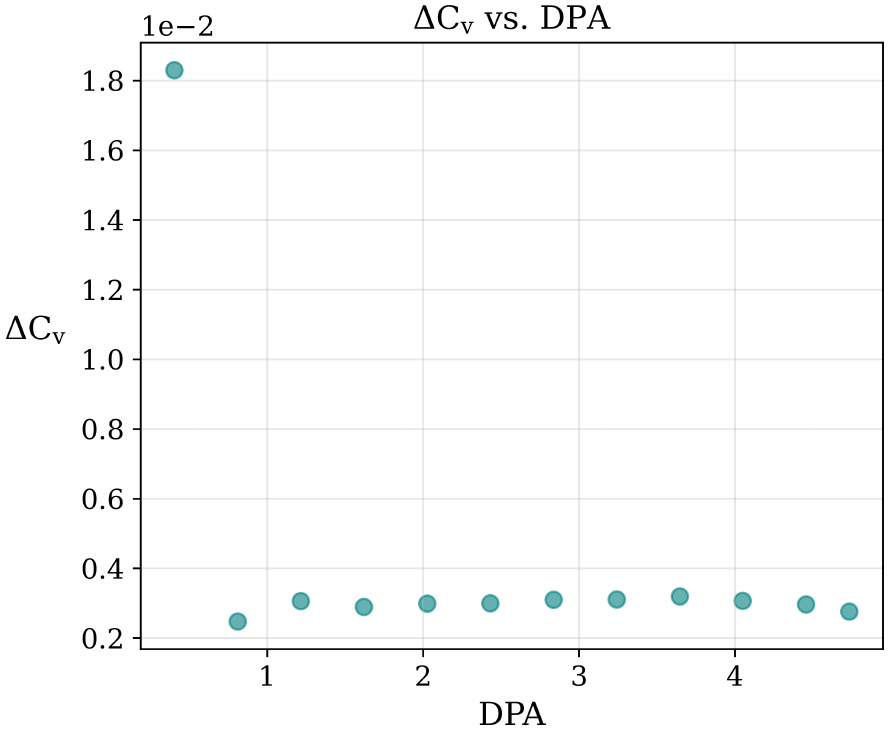}
    \caption{Vacancy concentration evolution with increasing dose for a periodic pulsed beam experiment. The initial large increase represents the change from only thermal point defects (here we deem this to be negligible) to a high concentration, while the increases at all subsequent doses are much smaller as the proliferation of radiation defects from initial irradiation quickly reduces later vacancy concentrations.}
    \label{fig:Delta_cv}
\end{figure}

Figure \ref{fig:Pulsed_beam} shows the evolution of the SAW frequency with time, identifying the periods where the beam is on (white background and SAW frequency values represented by orange dots) and off (gray background and SAW frequency represented by blue dots). For each of the beam on/off periods, the aforementioned fitting has been calculated, obtaining the increment in stable vacancies in the material with increasing dose (see Figure \ref{fig:Delta_cv}).

\begin{figure}[h]
    \centering
    \includegraphics[width=\textwidth]{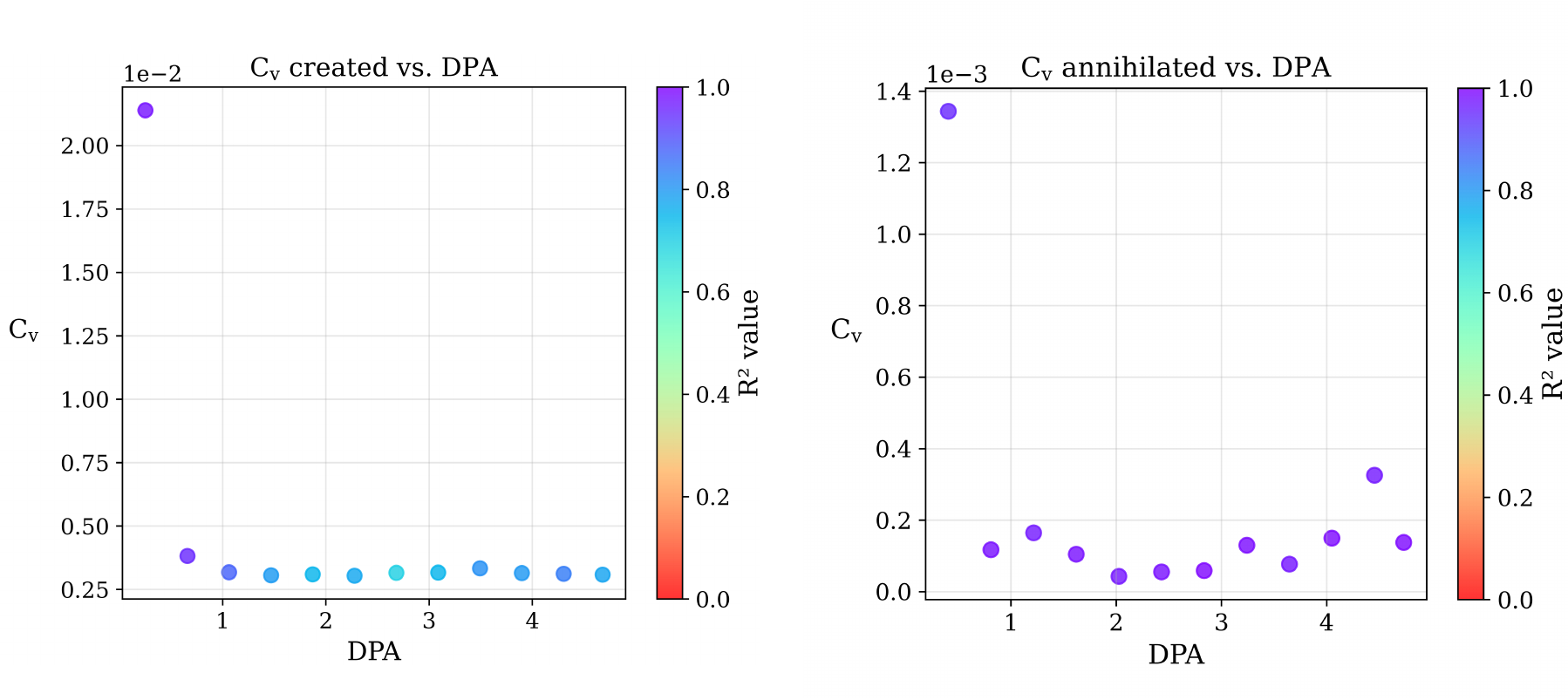}
    \caption{Vacancy concentration calculated for each beam pulse during the logarithmic pulsed-beam experiment (Fig. \ref{fig:Pulsed_beam}). The concentration of vacancies created corresponds to the beam on periods, and the concentration of vacancies annihilated corresponds to the beam off periods. The color of each point corresponds to the quality of the data, measured by $R^2$ value.}
    \label{fig:cv_r2}
\end{figure}

The quality of the estimated vacancy concentration for the beam-on and beam-off fits is presented on the right axis of Figure \ref{fig:cv_r2}, expressed through their corresponding $R^2$ values. As shown, the beam-off case generally yields higher $R^2$ values, consistent with the fact that an exponential model tends to accommodate data variability more readily than a square-root dependence. 

\subsection{Log-scale pulsed ion beam experiments}
\begin{figure}[h]
    \centering
    \includegraphics[width=\textwidth]{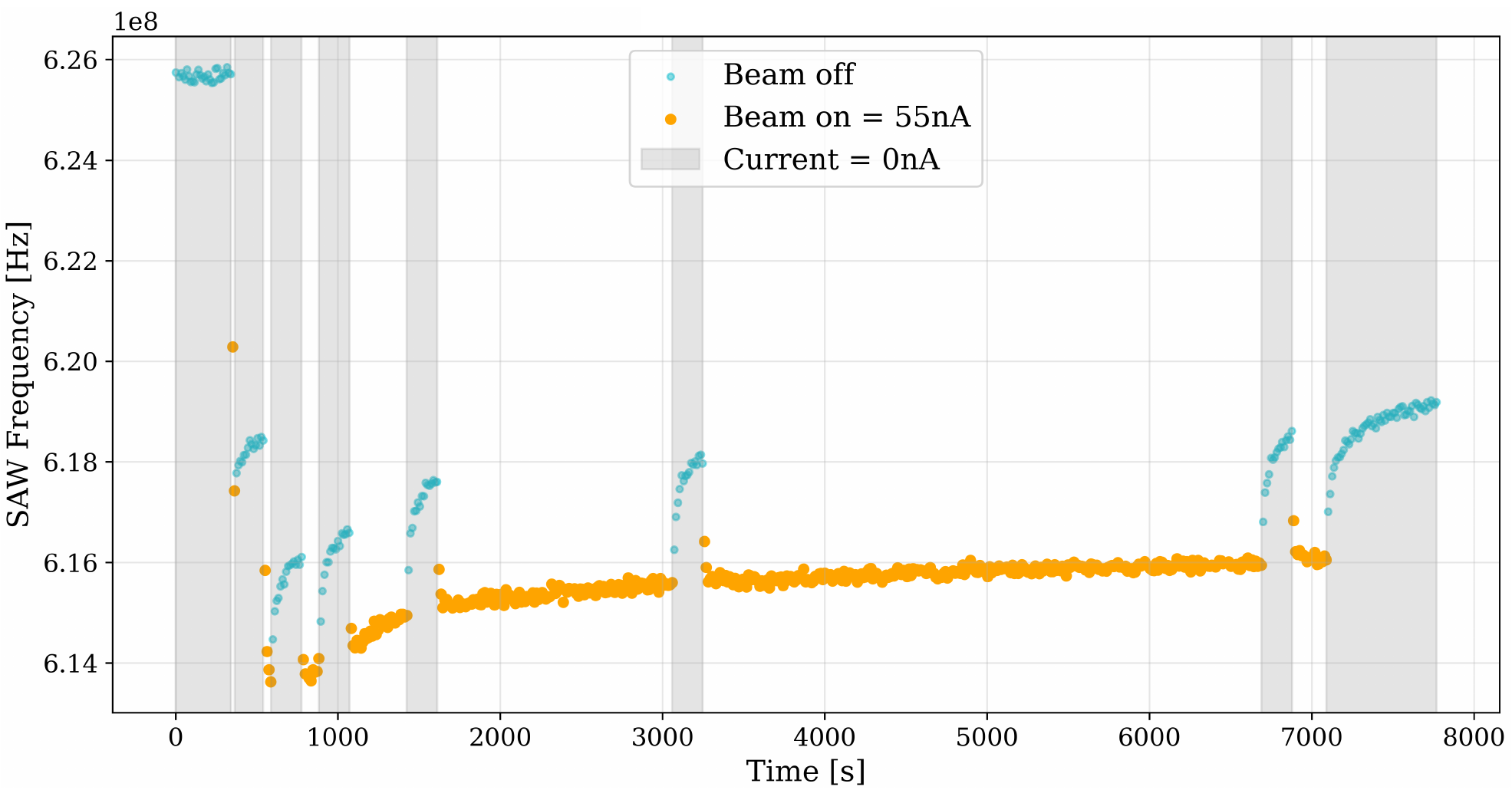}
    \caption{Logarithmic pulsed beam experiment. The white background and orange dots represent the periods in which the beam is on, and the grey background and blue dots represent the periods in which the beam is off.}
    \label{fig:Pulsed_log}
\end{figure}

\begin{figure}[h]
    \centering
    \includegraphics[width=\textwidth]{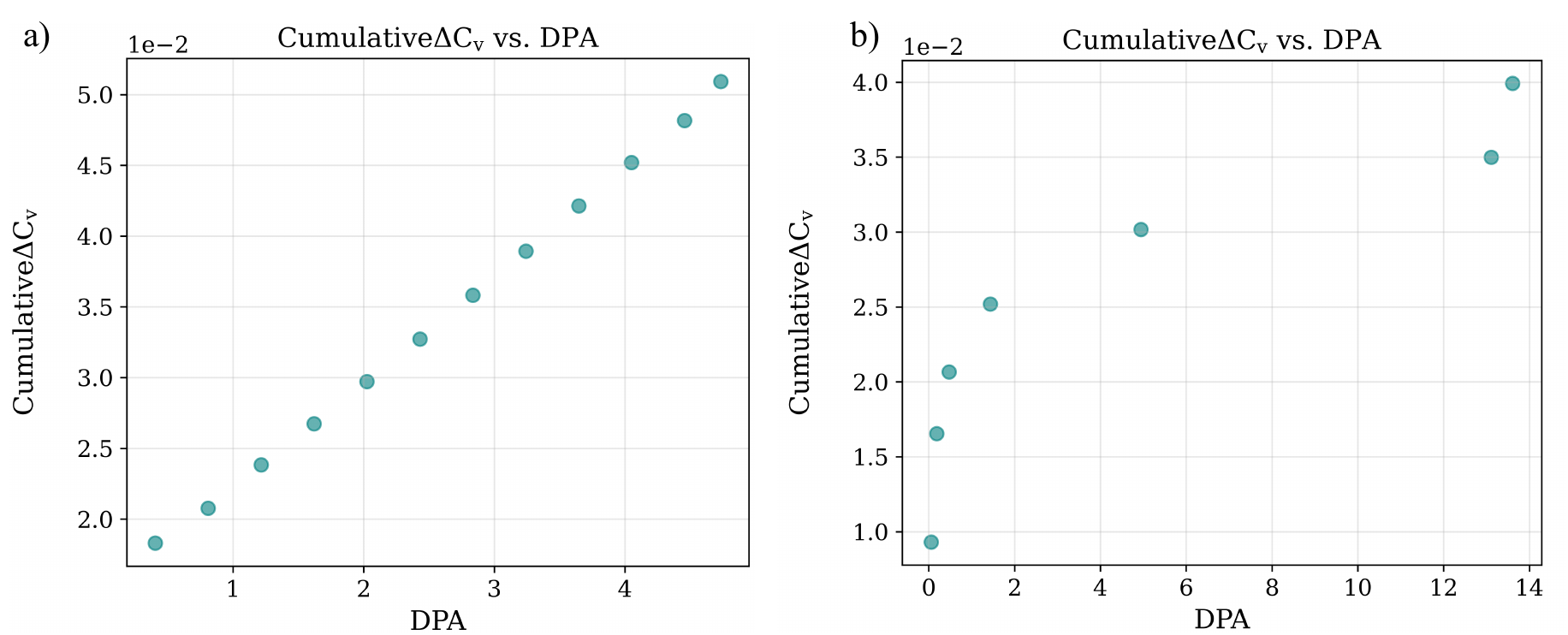}
    \caption{Cumulative vacancy concentration comparison between a) the periodic pulsed-beam, and b) the logarithmic scale pulsed-beam experiments.}
    \label{fig:Cum_cv}
\end{figure}

To monitor the evolution of the vacancy concentration after point defects have reached their steady-state concentration under each pulse of ion beam irradiation, allowing time for defects to evolve and form clusters (voids, dislocation loops, stacking-fault tetrahedra) or to annihilate through recombination or absorption at grain boundaries, the beam was switched on and off each time the sample accumulated approximately half-logarithmic increments of DPA (0.03, 0.1, 0.3, 1, 3, and 10 DPA). Figure \ref{fig:Pulsed_log} shows the results of this experiment for the same material irradiated during the periodic pulsed-beam experiment.
Interestingly, and consistent with defect-evolution dynamics, the cumulative increase in vacancy concentration exhibits an initial exponential rise, followed by a plateau, and then a second increase. This behavior contrasts with the response shown in Figure \ref{fig:Cum_cv}.b) for the periodic pulsed beam, where the cumulative vacancy concentration increases linearly.

\subsection{Vacancy concentration estimation comparison with TEM void quantification}
To demonstrate the accuracy of this methodology to estimate vacancy concentration during \emph{in situ} ion irradiation, the ratio of the total vacancy concentration between 2 different compositions has been compared with the amount of voids visualized by TEM. 

The alloys selected for this experiment have been two compositions from the CuCrTa system, as the downside of GRCop-84 for fusion applications is that Nb can barely be used in a material that is going to be exposed to the fusion neutron spectrum \cite{fetter_long-term_1990} as it becomes highly radioactive. Then, another candidate element from which Nb could be replaced is Ta, as this element also form C15 Laves phases with Cr \cite{kumar_structural_2000}, which are responsible for the good mechanical properties of the GRCop family. For this reason, one composition to be tested is similar to GRCop-84, but using Ta \cite{wallace_ultra-rapid_2022}, being Cu–9Cr–4Ta (at.\%). The other sample will be Cu–24Cr–5Ta (at.\%) in order to expect a clearly different behavior between them.

\begin{figure}[h]
    \centering
    \includegraphics[width=\textwidth]{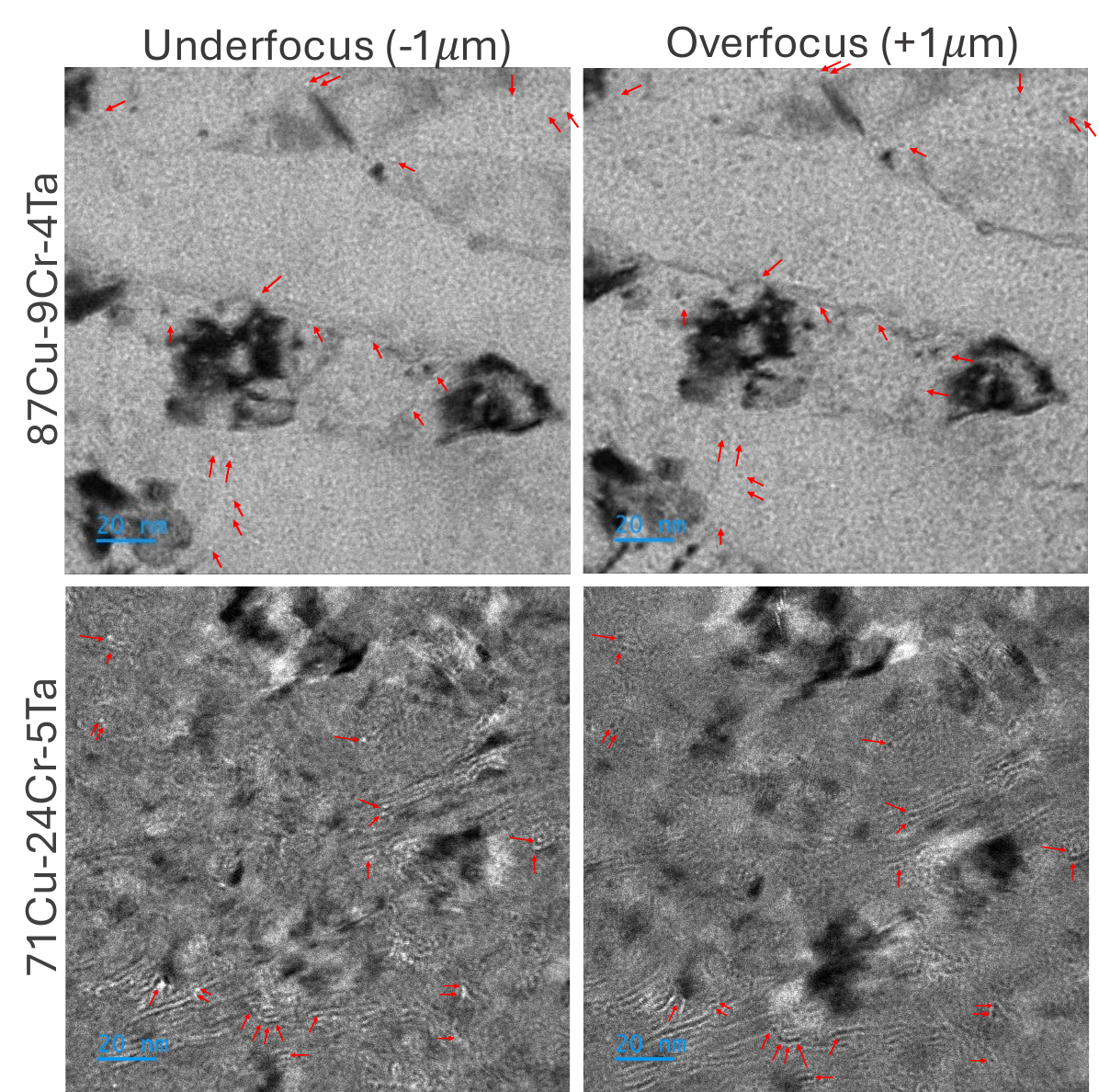}
    \caption{TEM images from the different Cu-Cr-Ta compositions. On the left side pictures were taken with an underfocus value of $-1\mu m$, and images on the right column were taken with an overfocus value of $+1\mu m$.}
    \label{fig:TEM}
\end{figure}

\begin{figure}[h]
    \centering
    \includegraphics[width=0.95\textwidth]{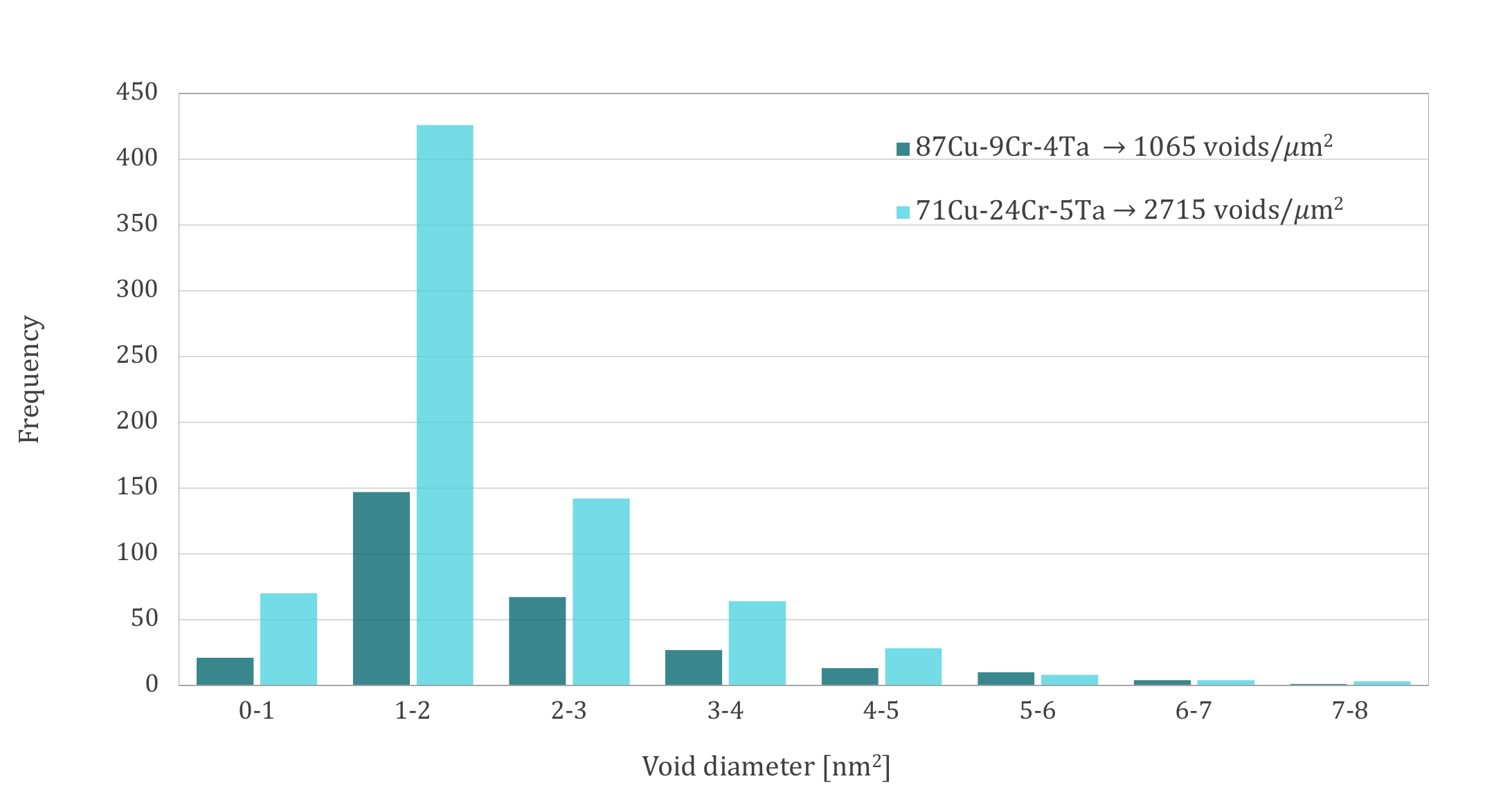}
    \caption{Distribution of the number and size of voids detected on each sample. Void densities per $\mu m ^2$ for each composition are shown in the upper right.}
    \label{fig:TEM_dist}
\end{figure}

Voids were detected in CTEM mode by the overfocus-underfocus technique \cite{williams_transmission_2008}, taking advantage of the Fresnel contrast, using an overfocus-underfocus increase of $\pm1 \mu m$. Figure \ref{fig:TEM} shows the result, with red arrows denoting the voids in the underfocus (a) and overfocus (b) situations. After looking into different images for each sample, counting the voids and the area examined using Fiji, the void density (voids per area [$nm^2$]) was calculated, as well as the distribution of void size and number per composition (see Figure \ref{fig:TEM_dist}).

\FloatBarrier
\section{Model development}
\label{sec:mod_dev}
\subsection{Cascade point defect evolution}
\label{sec:mod_dev_cas}
During radiation cascades, multiple Frenkel pairs are created, some of which will be annihilated during the ``annealing'' period of the radiation cascade, while others would remain in the microstructure contributing to the point defect concentration that, at the end, will be responsible for the degradation of the material's properties. The equilibrium concentration of each of these species is determined by recombination/annihilation and the creation of new point defects during consecutive radiation cascades. Then, the point defect balance equations in the presence of successive radiation cascades are:
\begin{equation}
\begin{aligned}
\frac{dC_v}{dt} = K_0(1-\epsilon_{r})(1-\epsilon_{v}) - K_{iv} C_i C_v - K_{vs} C_v C_s +L_{v}  \\
\frac{dC_i}{dt} = K_0(1-\epsilon_{r})(1-\epsilon_{i}) - K_{iv} C_i C_v - K_{is} C_i C_s
\end{aligned}
\label{eq:pointdefects_g}
\end{equation}
where the subscripts \emph{i}, \emph{v} and \emph{s} denote interstitials, vacancies, and sinks, respectively, $C [\#/m^3]$ is the concentration, $K_{0} [DPA/s]$ is the production rate, $K_{xy} [m^3/s]$ is the recombination rate coefficient for components denoted by the subscripts $x$ and $y$. $L_{v}[\#/m^3s]$ is the production of thermal vacancies from sinks, $\epsilon_{r}$ is the on-cascade vacancy-interstitial recombination rate and $\epsilon_{i}$ or $\epsilon_{v}$ is the fraction of interstitials/vacancies that will migrate to a pre-existing cluster of the same species \cite{was_fundamentals_2017}. Then, sink strengths are defined by:

\begin{equation}
K_{vs}=4 \pi r_{vs} D_v
\end{equation}

\begin{equation}
K_{is}=4 \pi r_{is} D_i
\end{equation}

\begin{equation}
K_{iv}=4 \pi r_{iv} (D_i + D_v)   
\end{equation}
but as $D_i >> D_v$,
\begin{equation}
    K_{iv} \approx 4 \pi r_{iv} D_i 
\end{equation}

In nanocrystalline materials, as is the case of the samples presented here, the cascade recombination rate between interstitials and vacancies is smaller than for  non-nanocrystalline materials, as mobile interstitials will find a grain boundary far more rapidly than a vacancy \cite{samaras_computer_2002}, so 0.1 has been selected for $\epsilon_{r}$, implying that around 10\% of the vacancies will recombine with interstitials. On the other hand, at low temperatures, vacancy clusters increase the energy of the system, being an unstable defect, so they tend to recombine or eject vacancies. Additionally, saturated grain boundaries will eject interstitials from them that will recombine with pre-existing defects, like staking-fault tetrahedral (SFT), voids, etc, before some vacancies have the opportunity to find them \cite{thomas_thermal_2023, bai_efficient_2010}. Considering all of this, at room temperature irradiations the fraction of vacancies that contribute to pre-existing vacancy clusters is negligible, making $\epsilon_{v}=0$. Finally, it is important to mention that at low temperatures ($T\leq 0.5T_m $, where $T_m$ is the melting temperature) the production of thermal vacancies from sink sources is not significant \cite{was_fundamentals_2017}, then $L_{v}=0$. Considering all these conditions, the expression for vacancy evolution in equation \eqref{eq:pointdefects_g} can be simplified to:
\begin{equation}
\frac{dC_v}{dt} = 0.9 K_0 - K_{iv} C_i C_v - K_{vs} C_v C_s 
\label{eq:pointdefects_v}
\end{equation}

\begin{figure}[h]
    \centering
    \includegraphics[width=0.85\textwidth]{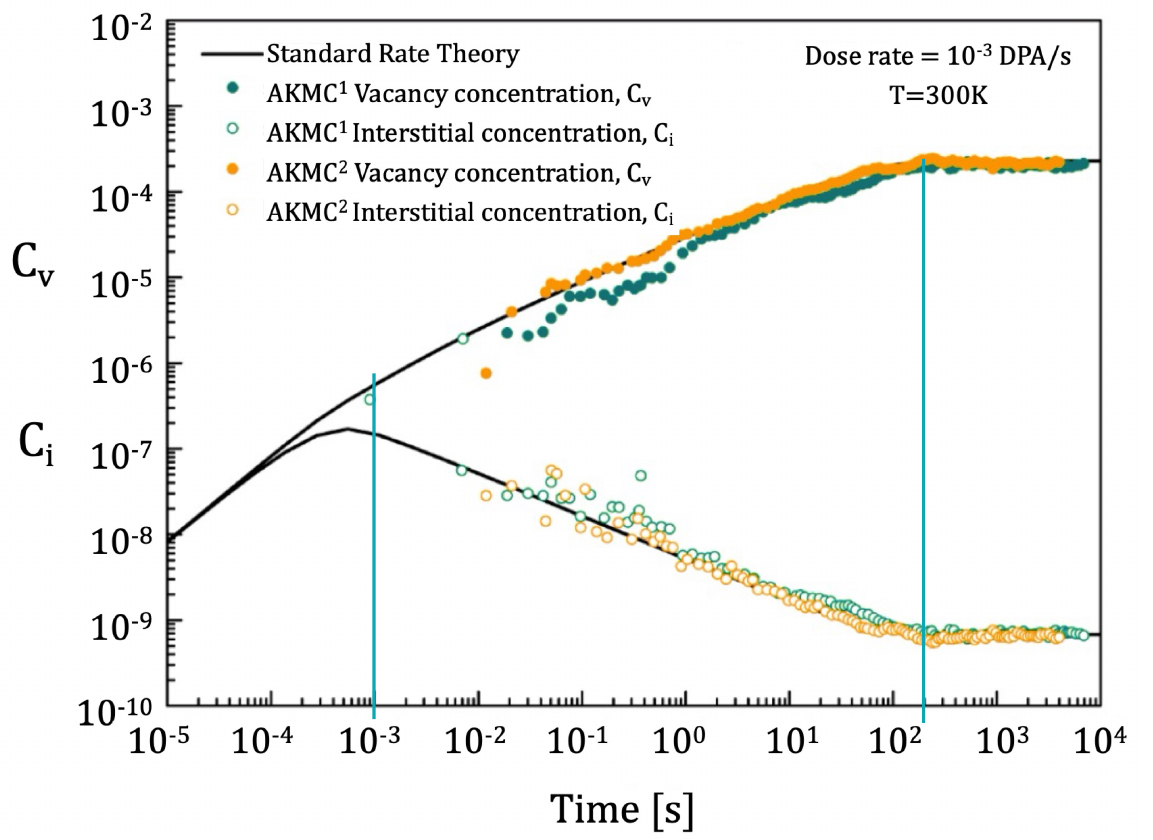}
    \caption{Point defect evolution for FeNi at 300K and 10-3 DPA/s, obtained from standard rate theory and atomistic kinetic Monte Carlo (AKMC) simulations using a random sink distribution and an annihilation rate of sinks. Plot adapted from \cite{martinez_point_2020}.}
    \label{fig:SRT_AKMC}
\end{figure}

The time frame in which the SAW frequency evolves and stabilizes when the beam is turned on and off in our experiments (see Figure \ref{fig:TGS_beam_comparison}.a), is similar to the time frame in which the vacancy/interstitial concentrations evolve, on the order of minutes \cite{martinez_point_2020} for the experimental conditions presented here, therefore we investigated the relation of the point defect evolution formulas with the SAW frequency.
Previous computational simulations from E. Martinez et al. \cite{martinez_point_2020}, employing standard rate theory (SRT) and atomistic kinetic Monte Carlo (akMC), have shown that the time frame in which the period of mutual recombination occurs during the equilibration process of point defects at the beginning of the irradiation, is in the range between 0.0005 - 200 seconds for an irradiation performed at 300K and using a dose rate of $10^{-3} DPA/s$ (see Figure \ref{fig:SRT_AKMC}), which matches the relaxation times that we observe in our experiments in similar conditions.

As seen in Figure \ref{fig:TGS_beam_comparison}, the evolution of the SAW frequency shows a different behavior during the beam onset and offset, implying that different but related mechanisms are taking place in the material by the beam fluctuation. For this reason, we evaluate point defect concentration evolution for these two cases independently. When the beam is turned on, point defects in the material are created by the impinging ion beam ($K_0$), some of them will recombine between them ($K_{iv}$), and others will annihilate at grain boundaries ($K_{vs}$). Then, equation \ref{eq:pointdefects_v} describes what happens with the insterstitial concentration by the beam onset for a high sink material irradiated at low temperatures (RT) \cite{was_fundamentals_2017}. By equating this expression, the vacancy concentration in the regime prior when point defect equilibrium is reached, where mutual recombination is greater than interstitial recombination at sinks, would be:
\begin{equation}
C_v = \bigg(\frac{0.9 K_0 K_{vs} C_s t}{K_{iv}}\bigg)^{\tfrac{1}{2}} = \bigg(\frac{0.9 K_0 4 \pi r_{is} D_i C_s t}{4 \pi r_{iv} (D_{v} + D_{i})}\bigg)^{\tfrac{1}{2}} \approx (0.9 K_0 C_s t)^{\frac{1}{2}}
\label{eq:vac_con_on}
\end{equation}

Later, when the beam is turned off, there will be no creation of point defects, so $K_0 = 0$, and also all interstitials will have migrated to the grain boundary, therefore there will be no recombination between vacancy and interstitials, $K_{iv} = 0$. After applying these conditions to equation \ref{eq:pointdefects_v}, the following expression can be derived:
\begin{align}
\ln \frac{C_v}{C_{v_0}} &= -K_{vs} C_s (t - t_0)
\end{align}

\begin{align}
C_v = C_{v_0} e^{-K_{vs} C_s t}
\end{align}

\subsection{Relation between SAW frequency and point defect concentration}
Interstitials migrate more rapidly to sinks than vacancies \cite{was_fundamentals_2017}; consequently, their steady-state concentration is significantly lower. As shown in Figure \ref{fig:SRT_AKMC}, where the simulated conditions \cite{martinez_point_2020} are comparable to those employed in the present experiments, the interstitial concentration is approximately five orders of magnitude lower than the vacancy concentration. Therefore, the contribution of interstitials can be considered negligible relative to that of vacancies, and vacancies may be regarded as the dominant point defects influencing the mechanical properties of the material under primary radiation damage. Accordingly, the Young's modulus can be rewritten as the Young's modulus of a porous material with a small fraction of spherical pores \cite{hasselman_effect_1964}:

\begin{equation}
E_{p} = E_{0}(1 - aP)
\end{equation}

where $a$ is a parameter that will depend on the Poisson's ratio of the material:
\begin{equation}
a = 3(9+5\nu_0)\frac{(1-\nu_0)}{2(7-5\nu_0)}
\end{equation}
and $P$ the volume fraction of porosity defined as:
\begin{equation}
P=\frac{V_{voids}}{V_{Total}}
\end{equation}

Additionally, we know that the rate of growth for the void volume under irradiation conditions is defined by the following equation \cite{was_fundamentals_2017}:

\begin{equation}
\frac{dV_{void}}{dt}=4\pi R\Omega[D_v(C_v-C_v^V)-D_iC_i]
\label{eq:void_growth_rate}
\end{equation}

where $R [m]$ is the void radius, $\Omega [m^3]$ is the atomic volume and $C_v^V [\#/m^3]$ is the vacancy concentration contributing to the formation of other voids. Moreover, as the concentration of interstitials is negligible in comparison to the vacancy concentration (see Supplementary Material), as shown on Figure \ref{fig:SRT_AKMC}, equation \ref{eq:void_growth_rate} can be rewritten as: 
\begin{equation}
\frac{dV_{void}}{dt}=4\pi R\Omega D_v(C_v-C_v^V)
\label{eq:void_growth_rate_reduced}
\end{equation}

Then we conclude that the volume fraction of porosity is proportional to the vacancy concentration:

\begin{equation}
P=\frac{V_{voids}}{V_{Total}}\propto C_v
\label{eq:void_growth}
\end{equation}
where $V_{voids} = \sum {n V_{void}}$ and $n$ is the number of voids.

The previous expression of the Young's modulus can be rewritten as:

\begin{equation}
 E_{p} \approx E_{0}(1 - aC_{v})
\end{equation}

And substituting this formula into equation \ref{eq:SAW_to_E}, the SAW frequency can be expressed in terms of the vacancy concentration as follows:

\begin{align}
f_{SAW} &= \frac{1}{\lambda}\sqrt{\frac{E_{0}(1-aP)}{\rho}}
  \;\; \approx \;\; \frac{1}{\lambda}\sqrt{\frac{E_{0}(1-aC_{v})}{\rho}} \notag \\
&\approx \frac{1}{\lambda}\sqrt{\frac{E_{0}}{\rho}}\sqrt{(1-aC_{v})}
  \;\; \approx \;\; f_{SAW_0}\sqrt{(1-aC_{v})}
\label{eq:SAW_Cv}
\end{align}

\subsubsection{Evolution of the SAW frequency by beam onset and offset}
\label{mod_dev_ev}
The evolution of the SAW frequency of a material subjected to radiation damage can be described by Equation \ref{eq:SAW_Cv}, where an explicit expression for the time-dependent vacancy concentration must be substituted. Considering the specific conditions corresponding to the beam-on case, and replacing $C_v$ with the expression governing the vacancy concentration evolution during beam irradiation using the assumptions from section \ref{sec:mod_dev_cas}, we obtain:

\begin{equation}
f_{SAW} \approx f_{SAW_0} \sqrt{1 - a (0.9K_0 C_s t)^{\frac{1}{2}}}
\end{equation}

For the beam-off case, substituting $C_v$ with the expression for the vacancy evolution concentration just after the beam is off, we arrive to:

\begin{equation}
f_{SAW} \approx f_{SAW_0} \sqrt{1 - a (C_{v_0} e^{-K_{vs} C_s t})}
\label{eq:sawf_off_fit}
\end{equation}

At this stage, to determine the vacancy concentration for each case, the experimental data must be fitted to the corresponding expressions, from which the respective values can be extracted.

\begin{figure}[h]
    \centering
    \includegraphics[width=\textwidth]{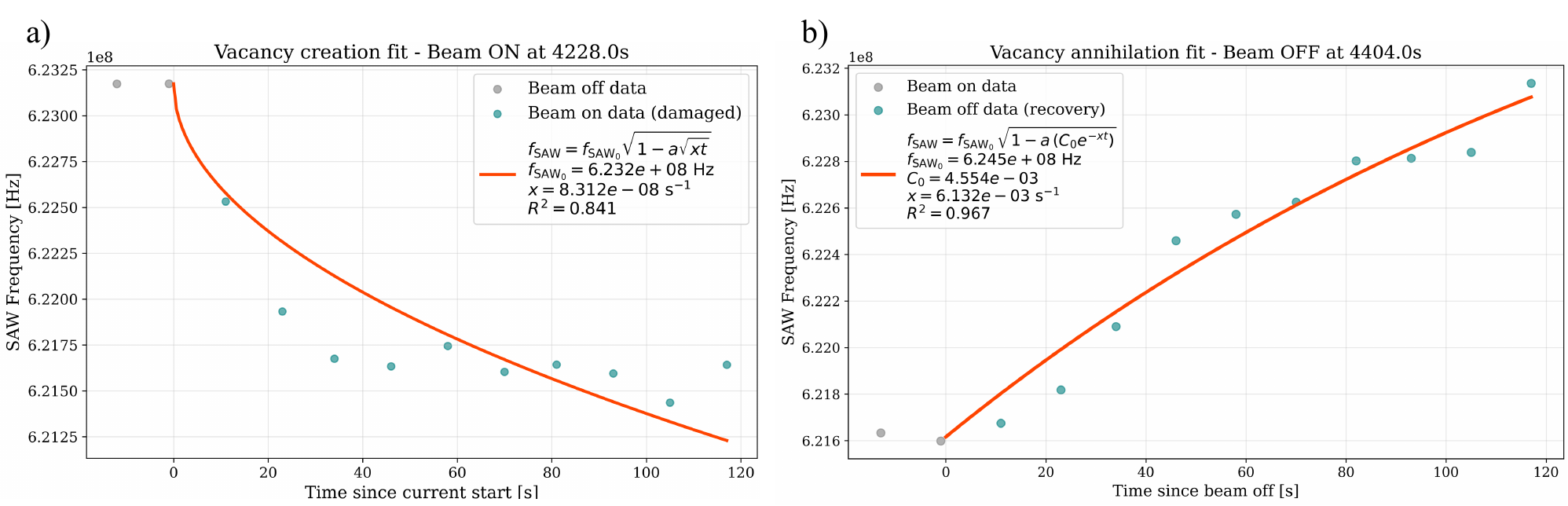}
    \caption{Beam on/off period fitting. The data from these fitting correspond to the $4228 s$ beam on period and the $4404 s$ beam off period from Figure \ref{fig:Pulsed_beam}.}
    \label{fig:Beamonoff_fit}
\end{figure}

The formula that will be fitted for the beam-on case has the following shape

\begin{equation}
y = y_0 \sqrt{1 - a (xt)^{\frac{1}{2}}}
\end{equation}

where $x$ would be the fitting constant equivalent to $0.9K_0C_s$. Therefore, $C_v$ can be obtained by substituting the value of $x$ in equation \ref{eq:vac_con_on}. For the beam-off case the formula to be fit will be of the form:

\begin{equation}
y=y_0 \sqrt{1 - a (b e^{-x t})}
\label{eq:beam_off_fit}
\end{equation}

where $b$ would be equivalent to $C_{v_0}$, the value of vacancy concentration obtained on the beam-on case of that beam pulse, which would be employed as an initial guess of $b$ for the fitting (see Figure \ref{fig:Beamonoff_fit}). In consequence, the concentration of vacancies that are annihilated after the beam is turned off can be estimated using by operating the variables inside the bracket in Equation \ref{eq:beam_off_fit}. For each fit, beam-on and beam off- cases, it goodness has been evaluated by the $R^2$ value.

Finally, by subtracting the number of annihilated vacancies from the number of vacancies generated, the remaining vacancy concentration in the material ($\Delta C_v$) can be determined as shown in Figure \ref{fig:Delta_cv}. This residual concentration serves as an indicator of the damage state induced by each pulse of the ion irradiation.

\FloatBarrier
\section{Discussion}
\subsection{Evolution of the TGS signal by beam onset}
After evaluating the behavior of thermoelastic properties of the CuCrNb system while they are being irradiated by Cu$^{3+}$ ions, an initial drop in both properties can be noticed in Figure \ref{fig:TGS_beam_comparison}.a) when the beam is first switched on, at approximately 700 s. Subsequently, the thermal diffusivity stabilizes near 5.75x10$^{-6}\ m^2s^{-1}$, exhibiting negligible variation during beam on/off cycling in comparison to the initial drop. In contrast, the SAW frequency increases following the initial decrease and continues to shift its trend in response to each beam modulation. Nabarro \cite{nabarro_effect_1952} calculated the effect that irradiation has on the elastic constants of cooper and sodium, showing that an increase of 1\% of vacant lattice sites would lead to a decrease of $\approx 2.3\%$ of the elastic constants. This fact is consistent with the observable fluctuations in the SAW frequency when the beam is turned on.
However, it should be noted that these kinetics occur concurrently with the transient temperature decrease and increase associated with beam shutdown and reactivation. As a result, the relative contributions of vacancy dynamics and temperature variations must be further analyzed.

\subsection{Beam temperature influence on TGS signal}
In Figure \ref{fig:TGS_beam_comparison}.a) it can be observed that the percentage change in the thermal diffusivity by the time the beam is turned on is around $-1.65\%$. Now, comparing this value with the thermal diffusivity vs. temperature plot from reference \cite{ellis_thermophysical_2000}, at room temperature, a change in thermal diffusivity of $-1.65\%$ would imply an increase in temperature of about $\approx 300^\circ K$, an unfeasible increase for the experimental conditions observed. 

Thermal infrared videos were recorded while the sample was being irradiated by Cu$^{3+}$ ions under different current and ionic charge conditions, being able to detect the real temperature increase in the sample produced by the beam. Therefore, for the case depicted in Figure \ref{fig:TGS_beam_comparison}.a), which corresponds to a beam current of 10 nA, the temperature in the sample increases by $\approx 1.5^\circ C$ according to Figure \ref{fig:TGS_beam_comparison}.d), which doesn't correspond with the expected temperature increase previously mentioned of $\approx 140^\circ C$. Consequently, this experiment indicates that the beam heating temperature effect cannot be responsible for the changes in thermo-elastic properties detected by the $I^3TGS$. Instead, another mechanism is causing the signal to fluctuate.

\subsection{Radiation-induced atom displacement captured by TGS}
A key distinction between heavy ion and proton irradiation lies in the differing nature of their interaction with the crystal lattice: heavy ions produce significant lattice disturbance through collision cascades, whereas protons, due to their much smaller effective cross section and stopping power, cause minimal structural disruption on a per-energy basis. Building on this principle, the experiment shown in Figure \ref{fig:TGS_beam_comparison} was designed to equalize the beam power for heavy ion and proton irradiation, ensuring comparable heat transfer to the sample while limiting lattice distortion in the proton case. By maintaining equivalent thermal input across both conditions, it becomes possible to distinguish whether changes in the SAW frequency and thermal diffusivity upon beam on/off cycling originate from radiation-induced microstructural modifications or from beam-induced heating effects. To calculate the beam conditions for the proton irradiation, while matching the same beam heating power than the one used during the Cu$^{3+}$ irradiation, the following formula was employed \cite{wiedemann_particle_2015}:
\begin{equation}
P = (e+q) U \frac{I}{q}
\end{equation}
being $U$ the terminal potential, $q$ the charge state, $e$ the elementary charge, and $I$ the ion beam current.
In Figure \ref{fig:TGS_beam_comparison}.b) the evolution of SAW frequency and thermal diffusivity with increasing dose under 1.624 MeV H$^{+}$ irradiation for the GRCop-84 sample is shown. In contrast with the heavy ion irradiation (Figure \ref{fig:TGS_beam_comparison}.a)), when using protons and matching the same beam heating power for both types of irradiations, there is almost no change in the SAW frequency, nor in the thermal diffusivity, indicating that any possible temperature influence from the ion beam has a negligible impact on the measurements. Hence, the changes observed in the TGS signals during irradiation are indeed due to radiation damage.

Additionally, since the thermal diffusivity can be influenced by the electronic stopping power during irradiation \cite{zarkadoula_effects_2019}, we calculated the absolute decrease in thermal diffusivity observed upon beam activation for H$^{+}$ irradiation and normalized it to the corresponding value for Cu$^{3+}$ irradiation. The resulting ratio, 11.2\%, is comparable in order of magnitude to the theoretical ratio of electronic stopping powers for protons and copper ions incident on copper, which is 19.6\% (see Supplemental Material for more details). While thermal diffusivity degradation is not expected to scale directly with the electronic stopping power \cite{saether_phonon_2022,momenzadeh_phonon_2013, zarkadoula_effects_2019}, the similarity between these values suggests that the fluctuations observed in the I$^{3}$TGS signal during irradiation are in large part governed by radiation damage mechanisms, consistent with our previous conclusion.

\subsection{Damage evolution monitoring based on beam onset and offset response}
After establishing that variations in the SAW frequency during beam onset and offset arise from radiation damage, we can confidently correlate these changes in the elastic response of the material with the vacancy concentration generated during irradiation, following the procedure described in Section \ref{sec:mod_dev}. 

Two different types of experiments were performed, with the main difference that in one of them the beam fluctuations were restricted to match the beam-on time with the vacancy concentration build-up period (around 2-3 min according to Figure \ref{fig:SRT_AKMC}), to restrict point defect clustering and/or evolution. However, in the other experiment the beam remained on until half-logarithmic values of DPA were reached, thereby allowing point defects, during some intervals, enough time to diffuse and reorganize. This hypothesis was corroborated by the evolution of the cumulative $\Delta C_v$ with increasing DPA shown in Figure \ref{fig:Cum_cv}.b), where a plateau appears during the intervals in which longer on-times were introduced. Moreover, when comparing both experiments up to 4 DPA, the periodically pulsed beam produces a linear increase in cumulative vacancy concentration, whereas the logarithmically pulsed beam results in an exponential trend, indicating that additional defect processes are taking place within the material.

\begin{figure}[h]
    \centering
    \includegraphics[width=\textwidth]{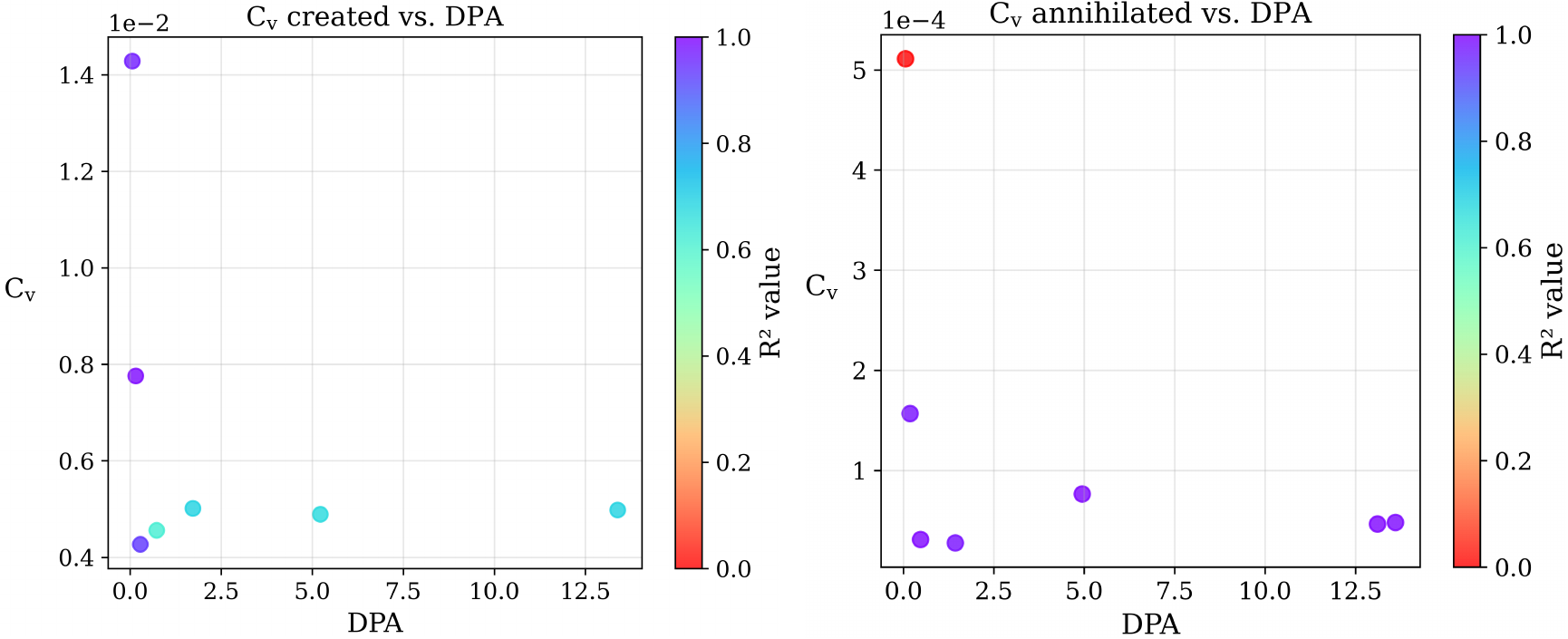}
    \caption{Vacancy concentration calculated for each beam pulse during the logarithmic pulsed-beam experiment (Figure \ref{fig:Pulsed_log}). the concentration of vacancies created correspond to the beam on periods, and the concentration of vacancies annihilated correspond to the beam off period. The color of each point corresponsd to the quality of the data, measured by its $R^2$ value.}
    \label{fig:cv_r2_log}
\end{figure}

Another notable observation is that the damage level in the same material is higher during the periodic pulse beam experiment. This is reasonable, as a larger fraction of free mobile vacancies remain available to contribute to fluctuations in the SAW frequency, unlike in the logarithmic pulsed-beam case, where many of the newly created vacancies have sufficient time to cluster, diffuse to grain boundaries, or recombine. Consequently, in the periodic pulsed-beam condition, one would expect a more uniform distribution of mobile vacancies that are poised to coalesce but have not yet done so. In contrast, during the logarithmic pulsed-beam experiment, the extended intervals allow a portion of the vacancies to annihilate or aggregate, reducing the number of free mobile vacancies remaining in the current pulses. This can be observed by comparing Figure \ref{fig:cv_r2} with Figure \ref{fig:cv_r2_log}, where the $C_v$ created in both cases is comparable, but the amount of $C_v$ annihilated is one order of magnitude smaller in the logarithmic experiment.

The last data point from Figure \ref{fig:Cum_cv}. b) shows an slight sudden increase, showing a noticeable increase of the cumulative vacancy concentration during the last shorter pulse, supporting the idea developed in the previous paragraph.

\subsection{Radiation damage resistance assessment}
\begin{figure}[h]
    \centering
    \includegraphics[width=\textwidth]{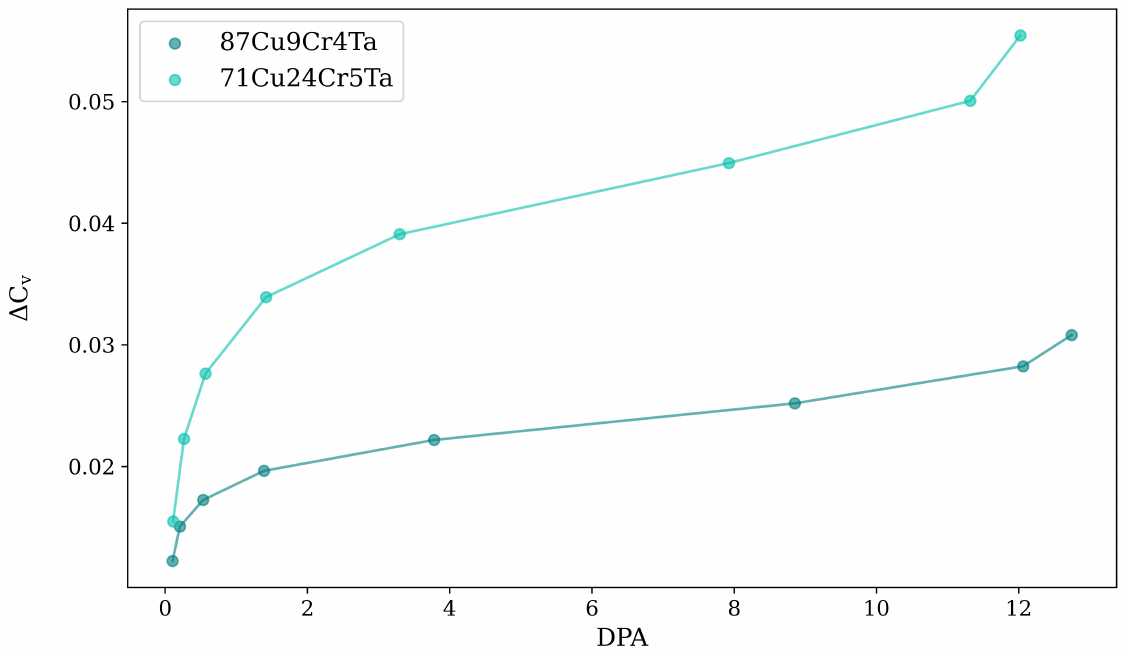}
    \caption{Cumulative vacancy concentration calculated for Cu-9Cr–4Ta (dark blue) and Cu-24Cr–5Ta (light blue) with increasing DPA, showing that the former, despite lower alloying element concentration, exhibits less vacancy concentration increase at similar doses.}
    \label{fig:cum_com}
\end{figure}

The cumulative vacancy concentration can serve as a comparative metric for assessing the radiation-damage resistance of materials manufactured under identical processing conditions, thereby minimizing microstructural variability that could otherwise influence the elastic response. In this work, two alloys from the CuCrTa system, Cu-9 at.\% Cr–4 at.\% Ta and Cu-24 at.\% Cr–5 at.\% Ta, were evaluated.
Figure \ref{fig:cum_com} shows the cumulative vacancy concentration for both alloys during the logarithmic pulsed-beam experiment. The results indicate that the Cu-9 at.\% Cr–4 at.\% Ta alloy exhibits superior resistance to radiation damage, as evidenced by its lower cumulative vacancy concentration at 12 DPA. This trend is consistent with the TEM analysis in Figure \ref{fig:TEM_dist}, which shows a smaller void density after 25 DPA in the Cu-9 at.\% Cr–4 at.\% Ta sample.
Although the DPA levels of the logarithmic pulsed-beam irradiations and the TEM measurements do not coincide, preventing a direct comparison between the two methods, an approximate correlation was established by linearly fitting the cumulative vacancy concentration versus DPA for each alloy and extrapolating to 25 DPA. The ratio of the predicted cumulative vacancy concentrations (Cu-9 at.\% Cr–4 at.\% Ta divided by Cu-24 at.\% Cr–5 at.\% Ta) is 0.5, which closely matches the ratio of void densities obtained from TEM (0.4). The agreement between both trends, in addition whit the other results shows on this paper,  highlights the reliability of TGS as a technique for estimating relative radiation-damage resistance under the conditions studied. Most intriguing is that the alloy with lower alloying element concentrations exhibits less than one half the vacancy buildup rate as measured by TGS and TEM, suggesting that chemical complexity alone cannot account for radiation damage resistance. This contradicts many recent suggestions that increased alloy entropy leads to more radiation damage resistance - this counterexample provides significant food for thought.

\FloatBarrier
\section{Conclusion}
In this work, we have demonstrated the potential of \emph{in situ} ion irradiation coupled with transient grating spectroscopy (I$^3$TGS) to monitor the real-time evolution of vacancy concentration generated by radiation damage in Cu-based alloys. By coupling beam onset and offset periods with temporal resolution TGS measurement, the dynamic response of SAW frequency to beam pulsing was captured, reporting reversible changes correlated with vacancy generation and annihilation kinetics.

Comparative irradiations using Cu$^{3+}$ and H$^{+}$ ions under equivalent beam heating conditions confirmed that the observed SAW frequency variations are primarily radiation-induced rather than thermally driven. Complementary infrared thermography analyses showed that the beam-induced temperature increase ($\approx 1.5^\circ C$) doesn't match the expected change in thermal diffusivity for GRCop-84, indicating that is not a temperature effect but a radiation damage effect. Additionally, comparison of the ratios of the electronic stopping power and the drop in thermal diffusivity between the Cu$^{3+}$ and H$^{+}$ ions, indicate that the phenomena governing fluctuations of the TGS signal are due to radiation.

The observed drops and recoveries of the SAW frequency following beam pulses are consistent with the vacancy concentration saturation predicted by standard rate theory and atomistic kinetic Monte Carlo simulation with the same temperature and dose rate conditions. The experimental time scales for equilibration of the SAW frequency fluctuations are comparably similar with simulated defect recombination kinetics, validating the use of TGS as a non-contact, non-destructive tool for real-time defect monitoring under irradiation. All of these results underscore the sensitivity of I$^3$TGS to variations in defect populations, allowing direct linkage between point defect evolution and macroscopic mechanical response.

In summary, experiments performed on this paper demonstrate that variations in the SAW frequency during beam onset and offset reliably reflect radiation-induced vacancy formation, enabling the use of cumulative vacancy concentration as an estimation of radiation damage. The comparison between periodically pulsed and half-logarithmically pulsed beams highlights the role of defect dynamics. Periodic pulsing maintains a higher fraction of free, mobile vacancies, leading to a more uniform distribution and linear accumulation, whereas logarithmic pulsing allows defect diffusion, clustering, and annihilation, resulting in exponential trends and slower vacancy accumulation. Applying this methodology to two CuCrTa alloys, candidate materials for RF antennas in fusion environments, shows that the Cu-9 at.\% Cr–4 at.\% Ta exhibits superior radiation-damage resistance despite its significantly lower alloying element concentration, consistent with both cumulative vacancy measurements and TEM observations of void density. The agreement between TGS predictions and TEM observations demonstrates that TGS is an effective technique for assessing relative radiation-damage resistance and capturing the influence of defect dynamics.

Future work will focus on expanding this technique in other materials systems, and employing it as a tool to perform material down-selection for fusion environments, helping to speed up solutions to the existing materials challenge.


\section*{Data availability statement}
The data and code that support the findings of this study are openly available in the following repository: https://github.com/shortlab/2025-ebotica-GRCop-Beam-Pulsing.

\section*{Declaration of generative AI and AI-assisted technologies in the manuscript preparation process}
During the initial preparation of this work, ChatGPT-5 was used by the lead author solely for English correction and expression. After using this tool, all authors reviewed and edited the content as needed, and take full responsibility for the content of the publication.  

\section*{CRedit authorship contribution statement}
\textbf{Elena Botica-Artalejo} Conceptualization, Methodology, Formal Analysis, Investigation, Writing – Original Draft, Writing – Review \& Editing, Visualization.  \textbf{Gregory M. Wallace} Conceptualization, Methodology, Supervision, Funding acquisition, Writing - Review \& Editing. \textbf{Michael P. Short} Conceptualization, Methodology, Supervision, Writing - Review \& Editing.

\section*{Declaration of Competing Interest} 
The authors declare that they have no known competing financial interests or personal relationships that could have appeared to influence the work reported in this paper.

\section*{Funding sources}
This work was supported by the U.S. Department of Energy, Office of Science, Office of  Fusion Energy Sciences under Award Number DE-SC0024307.

\section*{Acknowledgment}
This work was performed in part in the MIT.nano Characterization Facilities and in the Center of Nanoscale Systems at Harvard University. The authors gratefully acknowledge Florian Chavagnat for assistance with the IR camera setup, and Kevin Woller and Angus Wylie for their support in operating the ion accelerator.

\nocite{sivak_diffusion_2022}



\printbibliography
\end{document}